\documentclass{article}
            \usepackage{amsmath,amssymb}
            \usepackage{graphics}
            \begin{document}
           \title{Effects on quantum physics of the local availability of mathematics and space time dependent scaling factors for number systems}
            \author{Paul Benioff\\
            Physics Division, Argonne National
            Laboratory,\\ Argonne, IL 60439, USA\\
            e-mail: pbenioff@anl.gov}

            \maketitle
            
            \section{Introduction}
 The relation of mathematics to physics and its influence on physics have been a topic of much interest for some time.  A sampling of the  literature in this area includes Wigner's paper, "The Unreasonable Effectiveness of Mathematics in the Natural Sciences" \cite{Wigner} and many others \cite{Bernal,Bendaniel,Davies,Hut,Jannes,Omnes,Tegmark,Welch}. The approach taken by this author is to work towards a comprehensive theory of physics and mathematics together \cite{BenTCTPMTEC,BenTCTPM}.  Such a theory, if it exists, should treat physics and mathematics as a coherent whole and not as two separate but closely related entities.

   In this paper an approach is taken which may represent  definite steps toward such a coherent theory.  Two ideas form the base of this approach:  The local availability of mathematics and the freedom to choose scaling factors for number systems. Local availability of mathematics is based on the idea that all mathematics that an observer, $O_{x}$, at space time point, $x$, can, in principle, know or be aware of, is available locally at point $x$. Biology comes in to the extent that this locally available knowledge must reside in an observers brain. Details of how this is done, biologically, are left to others to determine.

   This leads to the association of a mathematical universe, $\bigvee_{x},$ to each point $x.$ $\bigvee_{x}$ contains all the mathematics that $O_{x}$ can know or be aware of.   For example, $\bigvee_{x}$ contains the various types of numbers:  the natural numbers, $\bar{N}_{x}$, the integers, $\bar{I}_{x}$, the rational numbers, $\overline{Ra}_{x},$  the real numbers, $\bar{R}_{x}$, and  the complex numbers, $\bar{C}_{x}$.  It also contains vector spaces, $\bar{V}_{x},$ such as  Hilbert spaces, $\bar{H}_{x},$ operator algebras, $\overline{Op}_{x},$ and many other structures.

   The universes are all equivalent in that any mathematical system present in one universe is present in another. It follows that  $\bigvee_{y}$ contains systems for the different types of numbers as $\bar{N}_{y},\bar{I}_{y}, \overline{Ra}_{y},\bar{R}_{y},\bar{C}_{y}.$ It also contains $\bar{H}_{y},\overline{Op}_{y}$, etc. Universe equivalence means here that for any system type,  $S$, $\bar{S}_{y}$ is the same system in $\bigvee_{y}$ as $\bar{S}_{x}$ is in $\bigvee_{x}.$

   For the purposes of this work, it is useful to have a specific definition of mathematical systems.  Here the mathematical logical definition of a system of a given type as a structure \cite{Barwise,Keisler} is used. A structure  consists of a base set, a few basic operations, none or a few basic relations, and a few constants. The structure must satisfy a set of axioms appropriate for the type of system being considered. For example, \begin{equation}\label{barNx}\bar{N}_{x}= \{N_{x},+_{x},\times_{x},<_{x},0_{x},1_{x}\}\end{equation} satisfies a set of axioms for the natural numbers as the nonnegative elements of a discrete
   ordered commutative ring with identity \cite{Kaye},
   \begin{equation}\label{barRx}\bar{R}_{x}=\{R_{x},+_{x},-_{x} ,\times_{x},\div_{x},<_{x}, 0_{x},1_{x}\}\end{equation} is a real number structure that satisfies the axioms for a complete ordered field \cite{real}, and \begin{equation}\label{barCx}\bar{C}_{x}=\{C_{x},+_{x}, -_{x},\times_{x},\div_{x}, 0_{x},1_{x}\}\end{equation} is a complex number structure that satisfies the axioms for an algebraically closed field of characteristic $0$  \cite{complex}. \begin{equation}\label{barHx}
   \bar{H}_{x}=\{H_{x},+_{x},-_{x},\cdot_{x},\langle-,-\rangle_{x},\psi_{x}\}
   \end{equation} is a  structure that satisfies the axioms for a Hilbert space \cite{Kadison}. Here $\psi_{x}$ is a  state variable in $\bar{H}_{x}.$ There are no constants in $\bar{H}_{x}.$ The subscript, $x,$ indicates that these structures are contained in $\bigvee_{x}.$

   The other idea introduced here is the use of scaling factors for  structures for the different number types. These scale structures are based on the observation \cite{BenIJTP,BenRENT,BenSPIE} that it is possible to define, for each number type, structures in which number values are scaled relative to those in the structures shown above. The scaling of number values must be compensated for by scaling of the basic operations and constants in such a way that the scaled structure satisfies the relevant set of axioms if and only if the original structure does.

    Scaling of number structures introduces scaling into other mathematical systems that are based on numbers as scalars for the system.  Hilbert spaces are examples as they are based on the complex numbers as scalars.

   The fact that number structures can be scaled allows one to introduce scaling factors that depend on space time or space and time. If $y=x+\hat{\mu}dx$ is a neighbor point of $x$, then the real scaling factor from $x$ to $y$ is defined by \begin{equation}\label{1stryx}
   r_{y,x}=e^{\vec{A}(x)\cdot\hat{\mu}dx}.\end{equation}Here $\vec{A}$ is a real valued gauge field that determines the amount of scaling, and $\hat{\mu}$ and $dx$ are, respectively, a unit vector and length of the vector from $x$ to $y.$ Also $\cdot$ denotes the scalar product.  For $y$ distant from $x,$ $r_{y,x}$ is obtained by a suitable path integral from $x$ to $y.$

   Space time scaling of numbers would seem to be problematic since it appears to imply that comparison of theoretical and experimental numbers obtained at different space time points have to be scaled to be compared.  This is not the case.  As will be seen,  number scaling plays no role in such comparisons.  More generally, it plays no role in what might be called, "the commerce of mathematics and physics".

    Space time dependent number scaling is limited to expressions in theoretical physics that require the mathematical comparison of mathematical entities at different space time points.  Typical examples are space time derivatives or integrals. Local availability of mathematics makes such a comparison problematic. If $f$ is a  space time functional that takes values in some structure $\bar{S},$ then "mathematics is local" requires that  for each point, $y$, $f(y)$ is an element of $\bar{S}_{y}.$ In this case space time integrals or derivatives of $f$ make no sense as they require addition or subtraction of values of $f$ in different structures. Addition and subtraction are defined only within structures, not between structures.

    This problem is solved by choosing some point $x$, such as an observers location, and transforming each $\bar{S}_{y}$ into a local representation of $\bar{S}_{y}$ on $\bar{S}_{x}.$ Two methods are available for doing this: parallel transformations for which the local representation of $\bar{S}_{y}$ on $\bar{S}_{x}$ is $\bar{S}_{x}$ itself, and  correspondence transformations. These give a local, scaled representation of $\bar{S}_{y}$ on $\bar{S}_{x}$ in that each  element of $\bar{S}_{y}$ corresponds to the same element of $\bar{S}_{x},$ multiplied by the factor $r_{y,x}.$

    The rest of this paper explains, in more detail, these ideas and some consequences for physics. The next  section describes representations of number types that differ by scaling factors. Sections \ref{FNS} and \ref{GF} describe space time fields of complex and real number structures and the representation of $r_{y,x}$ in terms of a gauge field, as in Eq. \ref{1stryx}. This is followed by a discussion of the local availability of mathematics and the assignment of separate mathematical universes to each space time point. Section \ref{CMPTM} describes correspondence and parallel transforms.  It is shown that $\vec{A}$ plays no role in the commerce of mathematics and physics. This involves the comparison and movement of the outcomes of theoretical predictions and experiments and the general use of numbers.

    Section \ref{QT} applies these ideas to quantum theory, both with and without the presence of $\vec{A}.$  Parallel and correspondence transformations are used  to describe the wave packet representation of a quantum system. It is seen that there is a wave packet description that closely follows what what is actually done in measuring the position distribution and position expectation value.  The coherence is unchanged in such a description.

    The next to last section uses "mathematics is local" and the scaling of numbers to insert $\vec{A}$ into gauge theories.  The discussion is brief as it has already been covered elsewhere \cite{BenIJTP,BenSPIE}. $\vec{A}$ appears in the Lagrangians as a boson for which a mass term is not forbidden. The last section concludes the paper.

    The origin of this work is based on aspects of mathematical locality that are already used in gauge theories \cite{Yang,Montvay} and their use in the standard model \cite{Novaes}. In these theories, an $n$ dimensional vector space, $\bar{V}_{x},$ is associated with each point, $x$, in space time. A matter field $\psi(x)$ takes values in $\bar{V}_{x}.$ Ordinary derivatives are replaced by covariant derivatives, $D_{\mu,x}$, because of the problem of comparing values of $\psi(x)$ with $\psi(y)$ and to introduce the freedom of choice of bases. These derivatives use elements of the gauge group, $U(n),$ and their representations in terms of generators of the Lie algebra, $u(n),$ to introduce gauge bosons into the theories.

 \section{Representations of different number types}\label{RDNT}

 Here the mathematical logical definition \cite{Barwise,Keisler} of mathematical systems as structures is used.  A structure consists of a base set, basic operations, relations, and constants that satisfy a set of axioms relevant to the system being considered. As each type of number is a mathematical system, this description leads to structure representations of each number type.

 The  types of numbers usually considered are the natural numbers, $\bar{N},$ the integers, $\bar{I},$ the rational numbers, $\overline{Ra},$ the real numbers, $\bar{R}$, and the complex numbers, $\bar{C}.$  Structures for the real and complex numbers can be defined by \begin{equation}\label{barRC}\begin{array}{c}\bar{R}= \{R,+,-,\times,\div,<,0,1\}\\\bar{C}=\{C,+,-,\times,\div,0,1\}.\end{array}
 \end{equation} A letter with an over line, such as $\bar{R}$, denotes a structure. A letter without an over line, as  $R$ in the definition of $\bar{R},$ denotes a base set of a structure.

 The main point of this section is to show, for each type of number, the existence of many structures that differ from one another by scale factors. To see how this works it is useful to consider a simple case for the natural numbers, $0,1,2,\cdots.$ Let $\bar{N}$ be represented by \begin{equation}\label{barN}\bar{N}=\{N,+,\times,<,0,1\}
 \end{equation} where $\bar{N}$ satisfies the axioms of arithmetic \cite{Kaye}.

 The structure $\bar{N}$ is a representation of the fact that $0,1,2\cdots$ with appropriate basic operations and relations are  natural numbers.  However,  subsets of $0,1,2,\cdots$, along with appropriate definitions of the basic operations, relations, and constants are also natural number structures.

 As an example, consider the even numbers, $0,2,4,\cdots$ in $\bar{N}$ Let  $\bar{N}_{2}$ be a structure for these numbers where \begin{equation}\label{barN2}
 \bar{N}_{2}=\{N_{2},+_{2},\times_{2},<_{2},0_{2},1_{2}\}\end{equation}  Here $N_{2}$ consists of the elements of $N$ with even number values in $\bar{N}.$ The structure $\bar{N}_{2}$ shows that the elements of $N$ that have value $2n$ in $\bar{N}$ have value $n$ in $\bar{N}_{2}.$ Thus the element that has value $2$ in $\bar{N}$ has value $1$ in $\bar{N}_{2},$ etc. The subscript $2$ on the constants, basic operations, and relations in $\bar{N}_{2}$ denotes the relation of these structure elements to those in $\bar{N}.$

 The definition of $\bar{N}_{2}$ floats in the sense that the specific relations of the basic operations, relation, and constants to  those in $\bar{N}$ must be specified. These are chosen so that $\bar{N}_{2}$ satisfies the axioms of arithmetic if and only if $\bar{N}$ does. A suitable choice that satisfies this requirement is another representation of $\bar{N}_{2}$ defined by \begin{equation}\label{barN22}
 \bar{N}^{2}_{2}=\{N_{2},+,\frac{\times}{2},<,0,2\}.
 \end{equation}This structure is called the representation of $\bar{N}_{2}$ on $\bar{N}.$

  $\bar{N}^{2}_{2}$ shows explicitly the relations between the basic operations, relations, and number values in $\bar{N}_{2}$ and and those in  $\bar{N}$. For example, $1_{2}\leftrightarrow 2,+_{2}\leftrightarrow +,\times_{2}\leftrightarrow \times /2,<_{2}\leftrightarrow <.$  These relations are such that $\bar{N}^{2}_{2},$  and thereby $\bar{N}_{2},$ satisfies the axioms of arithmetic if and only if $\bar{N}$ does.

  $\bar{N}^{2}_{2}$ also shows  the presence of $2$ as a scaling factor. Elements of the base set $N_{2}$ that have value $n$ in $\bar{N}_{2}$ have value $2n$ in $\bar{N}.$ Note that, by themselves, the elements of the base set have no intrinsic number values.  The values are determined by the axiomatic properties of the basic operations, relations, and constants in the structure containing them.

 This description of scaled representations applies to the other types of numbers as well. For real numbers let $r$ be a positive real number in $\bar{R}$, Eq. \ref{barRC}. Let \begin{equation}\label{barRr}
 \bar{R}_{r}=\{R,+_{r},-_{r},\times_{r},\div_{r},<_{r},0_{r},1_{r}\}
 \end{equation} be another real number structure. Define the representation of $\bar{R}_{r}$ on $\bar{R}$ by the structure, \begin{equation}\label{barRrr}
 \bar{R}^{r}_{r}=\{R,+,-,\frac{\times}{r},r\div,<,0,r\}. \end{equation} $\bar{R}^{r}_{r}$ shows that number values in $\bar{R}_{r}$ are related to those in $\bar{R}$ by a scaling factor $r$.

 $\bar{R}^{r}_{r}$ gives the definitions of the basic operations, relation, and constants in $\bar{R}_{r}$ in terms of those in $\bar{R}.$ These definitions must satisfy the requirement that $\bar{R}_{r}$ satisfies the real number axioms if and only if $\bar{R}^{r}_{r}$ does if and only if $\bar{R}$ does.\footnote{The relations between the structures are also valid for negative values of $r,$ provided $<$ in Eq. \ref{barRrr} is replaced by $>,$ and appropriate changes are made in the axioms to reflect this replacement.}

 Note that the base set $R$ is the same for all three structures.  Also the elements of $R$ do not have intrinsic number values independent of the structure containing $R.$  They attain number values  only inside a structure where the values  depend on the structure containing $R.$

  The relationships between number values in $\bar{R}^{r}_{r},\bar{R}_{r},$ and $\bar{R}$ can be represented by a new term, correspondence.  One says that the number value $a_{r}$ in $\bar{R}_{r}$ \emph{corresponds} to the number value $ra$ in $\bar{R}$.  This is different from the notion of \emph{sameness}.  In $\bar{R}$,  $ra$ is different from the value $a.$ However, $a$ is the \emph{same} value in $\bar{R}$ as $a_{r}$ is in $\bar{R}_{r}$ as $ra$ is in $\bar{R}^{r}_{r}.$ The distinctions between the  concepts of correspondence and sameness does not arise in the usual treatments of numbers.  The reason is that  sameness and correspondence coincide when $r=1.$

 For complex numbers, the structures, in addition to $\bar{C}$, Eq. \ref{barRC}, are\begin{equation}\label{barCc}\bar{C}_{r}=\{C,+_{r},-_{r},\times_{r}, \div_{r},0_{r},1_{r}\},\end{equation} and the representation of $\bar{C}_{r}$ on $\bar{C}$ as \begin{equation}\label{barCcc}
 \bar{C}^{r}_{r}=\{C,+,-,\frac{\times}{r},r\div,0,r\}.\end{equation} Here $r$ is a real number value in $\bar{C}.$  $a$ is the same number value in $\bar{C}$ as $a_{r}$ is in $\bar{C}_{r}.$ Otherwise the description is similar to that for the natural and real numbers.  More details on these and other number type representations  are given in \cite{BenRENT}.

 \section{Fields of mathematical structures}\label{FNS}
 As was noted in the introduction, the local availability of mathematics results in the assignment  of separate structures, $\bar{S}_{x},$ to each point, $x,$ of space time. Here $S$ denotes a type of mathematical structure. The discussion is limited to the main system types of concern. These are the real numbers, the complex numbers, and Hilbert spaces. Hilbert spaces are included here because the freedom of choice of scaling factors for number types affects Hilbert spaces as they are based on complex numbers as scalars.

 \subsection{Complex numbers}

  Parallel transformations between $\bar{C}_{x}$ and $\bar{C}_{y}$ for two points, $x,y,$ define the notion of \emph{same} number values between the structures. Let $F_{y,x}$ be an isomorphism from $\bar{C}_{x}$ onto $\bar{C}_{y}.$ With
 \begin{equation}\label{barCxCy}\begin{array}{c}
 \bar{C}_{x}=\{C_{x},+_{x},-_{x},\times_{x},\div_{x},0_{x},1_{x}\}\\ \bar{C}_{y}=\{C_{y},+_{y},-_{y},\times_{y},\div_{y},0_{y},1_{y}\},
 \end{array}\end{equation}and\begin{equation}\label{FyxCx}F_{y,x}\bar{C}_{x}= \bar{C}_{y},\end{equation} $F_{y,x}$ defines the notion of same number value and same operation in  $\bar{C}_{y}$ as that in $\bar{C}_{x}.$ This is expressed by \begin{equation}\label{ayFyxax}
 \begin{array}{c}a_{y}=F_{y,x}a_{x}\\[2mm]
 Op_{y}=F_{y,x}Op_{x}.\end{array}
 \end{equation} Here $a_{y}$ is the same (or $F_{y,x}$-same) number value in $\bar{C}_{y}$ as $a_{x}$ is in $\bar{C}_{x}.$ $Op_{y}$ is the same operation in $\bar{C}_{y}$ as $Op_{x}$ is in $\bar{C}_{x}.$ $Op$ denotes any one of the operations, $+,-,\times,\div.$

 Note that $F_{y,x}$ is independent of paths between $x$ and $y.$ This follows from the requirement that for a path P  from $x$ to $y$ and a path $Q$ from $y$ to $z$, \begin{equation}\label{FQPzx}F^{Q*P}_{z,x}=F^{Q}_{z,y}F^{P}_{y,x}.
 \end{equation} Here $Q*P$ is the concatenation of $Q$ to $P$.  If $z=x$  then the  path is cyclic and the final structure is identical to the initial one. This gives the result that \begin{equation}\label{FQPxx}
 F^{Q*P}_{x,x}=1.\end{equation} This shows that $F_{y,x}^{P}$ is path independent so that a path label is not needed. Note  that
 \begin{equation}\label{FyxFxy}F_{y,x}=F_{x,y}^{-1}.\end{equation} The subscript order in $F_{y,x}$ gives the path direction, from $x$ to $y.$

 At this point the freedom to choose complex number structures at each space time point is introduced.  This is an extension, to number structures, of the freedom  to choose basis sets in vector spaces as is used in gauge theories \cite{Montvay,Yang} This can be accounted for by factoring $F_{y,x}$ into a product of two isomorphisms as in \begin{equation}\label{FWW}
 F_{y,x}=W^{y}_{r}W^{r}_{x}.\end{equation} Here $y=x+\hat{\nu}dx$ is taken to be a neighbor point of $x.$

 The action of $W^{y}_{r}$ and $W^{r}_{y}$ is given by\begin{equation} \label{CyWWCx}\bar{C}_{y}=W^{y}_{r}\bar{C}^{r}_{x}=W^{y}_{r}W^{r}_{x} \bar{C}_{x}=F_{y,x}\bar{C}_{x}.\end{equation} Here $r_{y,x}$ is a real number in $\bar{C}_{x}$ that is associated with the link from $x$ to $y$. As was the case for $F_{y,x}$ the order of the subscripts determines the direction of the link.  Thus $r_{x,y}$ is a number in $\bar{C}_{y}$ for the same link but in the opposite direction and \begin{equation}\label{cyxcxy}(r_{x,y})_{x}r_{y,x}=1.
 \end{equation} Here $(r_{x,y})_{x}$ is the same number value in $\bar{C}_{x}$ as $r_{x,y}$ is in $\bar{C}_{y}.$ In the following, the subscripts $y,x$ are often suppressed on $r_{y,x}$ to simplify the notation.

 The structure $\bar{C}^{r}_{x}$ is defined to be the representation of $\bar{C}_{y}$ on $\bar{C}_{x}$. As is the case for $\bar{C}^{r}_{r},$ Eq. \ref{barCcc}, the number values and operations in $\bar{C}^{r}_{x}$ are defined in terms of the corresponding number values and operations in $\bar{C}_{x}$:

 \begin{equation}\label{barCcx}\bar{C}^{r}_{x}=\{C_{x},+_{x},-_{x}, \frac{\times_{x}}{r},r\div_{x},0_{x},r\}.\end{equation} The multiplication and division by $r,$ shown in $\times_{x}/r,r\div_{x},$  are operations in $\bar{C}_{x}.$ Note that the number value $r$ in $\bar{C}_{x}$  is the multiplicative identity in $\bar{C}^{r}_{x}.$ Also $\bar{C}^{r}_{x}$ has the same base set, $C_{x},$ as does $\bar{C}_{x}.$

 The corresponding definition of $W^{r}_{x}$ is given by \begin{equation} \label{Wcxdef}\begin{array}{c}W^{r}_{x}(a_{x})=ra_{x},\;\;\;\;  W^{r}_{x}(\pm_{x})=\pm_{x}\\\\ W^{r}_{x}(\times_{x})=\frac{ \times_{x}}{ r}\;\;\;\; W^{r}_{x}(\div_{x})=r\div_{x}.\end{array}\end{equation}$W^{r}_{x}$ is an isomorphism in that \begin{equation}\label{Wcxiso} W^{r}_{x}(a_{x}O_{x}b_{x}) =W^{r}_{x}(a_{x})W^{r}_{x} (O_{x})W^{r}_{x}(b_{x}).\end{equation} Here $a_{x}$ and $b_{x}$ are number values in $\bar{C}_{x}$ and $O_{x}$ denotes the basic operations in $\bar{C}_{x}.$

 $W^{y}_{r}$ has a similar definition as it is an isomorphism from $\bar{C}^{r}_{x}$ to $\bar{C}_{y}.$ Since the definition is  similar it will not be given here.

  $\bar{C}^{r}_{x}$ can also be represented in a form similar to that of Eq. \ref{barCc} as \begin{equation}\label{barCcx1}\bar{C}_{r,x}=\{C_{x}, \pm_{r,x},\times_{r,x} ,\div_{r,x},0_{r,x},1_{r,x}\}.
 \end{equation} This structure can be described as the  representation of $\bar{C}_{y}$ at $x.$ The relation between the number values and operations in $\bar{C}_{r,x}$ and those in $\bar{C}_{x}$  is provided by $\bar{C}^{r}_{x}$ which defines the number values and operations of $\bar{C}_{r,x}$ in terms of those in $\bar{C}_{x}.$  In this sense both $\bar{C}_{r,x}$ and $\bar{C}^{r}_{x}$ are different representations of the same structure.  From now on $\bar{C}_{r,x}$ and $\bar{C}^{r}_{x}$ will be referred to as the representation of $\bar{C}_{y}$ at $x$ and on $\bar{C}_{x}$ respectively.

 The relations between the basic operations and constants of  $\bar{C}^{r}_{x}$ and those of $\bar{C}_{x}$ lead to an interesting property.  Let $f^{r}_{x}(a^{r}_{x})$ be any analytic function on $\bar{C}^{r}_{x}.$ It follows that \begin{equation}\label{analf}f^{r}_{x}(a^{r}_{x})=b^{r}_{x} \Leftrightarrow r f_{x}(a_{x})=rb_{x}\Leftrightarrow f_{x}(a_{x})=b_{x}.
 \end{equation}Here $f_{x}$ is the same function on $\bar{C}_{x}$ as $f^{r}_{x}$ is on $\bar{C}^{r}_{x}.$ Also $a_{x}$ and $b_{x}$ are the same number values in $\bar{C}_{x}$ as $a^{r}_{x}$ and $b^{r}_{x}$ are in $\bar{C}^{r}_{x}.$

 This result follows from the observation that any term $(a^{r}_{x})^{n}/(b^{r}_{x})^{m}$ in $\bar{C}^{r}_{x}$ satisfies the relation  \begin{equation}\label{proddiv}\frac{(a^{r}_{x})^{n}}{(b^{r}_{x})^{m}} \mbox{}^{r}_{x}= r\frac{(a_{x})^{n}}{(b_{x})^{m}}\mbox{}_{x}.\end{equation} The $n$ factors and n-1 multiplications in the numerator contribute a factor of $r.$  This is canceled by a factor of $r$ in the denominator. The one $r$ factor arises from the relation of division in $\bar{C}^{r}_{x}$ to that in $\bar{C}_{x}.$

 Eq. \ref{analf} follows from the fact that Eq. \ref{proddiv} holds for each term in any convergent power series. As a result it holds for the power series itself.

 \subsection{Real numbers}
 Since the treatment for real numbers is similar to that for complex numbers, it will be  summarized here. The representations of $\bar{R}_{y}$ at $x$ and on $\bar{R}_{x}$ are given by Eqs. \ref{barRr} and \ref{barRrr} as
 \begin{equation}\label{RrxRrrx}\begin{array}{c}\bar{R}_{r,x}=\{R_{x},\pm_{r,x} \times_{r,x},\div_{r,x},<_{r,x}0_{r,x},1_{r,x}\}\\\bar{R}^{r}_{x}= \{R_{x},\pm_{x},\frac{\times_{x}}{r},r\div_{x},<_{x},0_{x},r_{x}\}.\end{array} \end{equation} Here $r=r_{x,y}$ is a positive real number.

 The definition of parallel transforms for complex number structures applies here also. Let $F_{y,x}$ transform $\bar{R}_{x}$ to $\bar{R}_{y}.$ $F_{y,x}$ defines the notion of same real number in that $a_{y}=F_{y,x}(a_{x})$ is the same real umber in $\bar{R}_{y}$ as $a_{x}$ is in $\bar{R}_{x}.$ As was shown in Eqs. \ref{FWW} and \ref{CyWWCx}, $F_{y,x}$ can be factored into two operators as in
 $F_{y,x}=W^{y}_{r}W^{r}_{x}$ where \begin{equation} \label{RyWWRx}\bar{R}_{y}=W^{y}_{r}\bar{R}^{r}_{x}= W^{y}_{r}W^{r}_{x} \bar{R}_{x}=F_{y,x}\bar{R}_{x}.\end{equation} $W^{r}_{x}$ defines the scaled representation of $\bar{R}_{y}$ on $\bar{R}_{x}.$ It is given explicitly by Eq. \ref{Wcxdef}.  $W^{y}_{r}$ maps the scaled representation onto $\bar{R}_{y}.$ Eqs. \ref{analf} and \ref{proddiv} also hold for the relations between any real valued analytic function on $\bar{R}^{r}_{x}$ and its correspondent on $\bar{R}_{x}$ in that $f^{r}_{x}(a^{r}_{x})=rf_{x}(a_{x}).$

  \subsection{Hilbert spaces}
  As noted in the introduction, Hilbert space structures have the form shown in Eq. \ref{barHx} as\begin{equation}\label{barHx1}
   \bar{H}_{x}=\{H_{x},+_{x},-_{x},\cdot_{x},\langle-,-\rangle_{x},\psi_{x}\}.
   \end{equation}  Complex numbers are included implicitly in that Hilbert spaces are closed under multiplication of vectors by complex numbers.  Also scalar products are bilinear maps with complex values.

   As was the case for numbers, parallel transformation of $\bar{H}_{y}$ to $x$ maps $\bar{H}_{y}$ onto $\bar{H}_{x}.$ If scaling of the numbers is included, then the local representation of $\bar{H}_{y},\bar{C}_{y}$ on $\bar{H}_{x},\bar{C}_{x}$ is given by $\bar{H}^{r}_{x},\bar{C}^{r}_{x}.$ The structure, $\bar{C}^{r}_{x},$ is shown in Eq. \ref{barCcx}. $\bar{H}^{r}_{x}$ is given by  \begin{equation}\label{barHrx}\bar{H}^{r}_{x}= \{H_{x},\pm_{x},\frac{\cdot_{x}}{r}, \frac{\langle -,-\rangle_{x}}{r},r\psi_{x}\}\end{equation}
   This equation gives explicitly the relations of  operations and vectors of the local representation of $\bar{H}_{y}$  to those in $\bar{H}_{x}.$ The relations are defined by the requirement that $\bar{H}^{r}_{x}$ satisfy the Hilbert space axioms \cite{Kadison} if and only if $\bar{H}_{x}$ does.\footnote{Support for the inclusion of $r$ as a vector multiplier, as in $\bar{H}^{r}_{x},$ Eq. \ref{barHrx}, is based on the equivalence between finite dimensional vector spaces and products of complex number fields \cite{Kadison}.  If $\bar{H}_{y}$ and $\bar{H}_{x}$ are $n$ dimensional spaces, then $\bar{H}_{y}\simeq \bar{C}_{y}^{n}$ and $\bar{H}_{x}\simeq\bar{C}_{x}^{n}.$

 These equivalences extend to the local representation of $\bar{H}_{y}$ on $\bar{H}_{x}.$ As the local representation of $\bar{C}_{y}$ on $\bar{C}_{x}$, $\bar{C}^{r}_{x}$ is the scalar field base for the local Hilbert space representation. It follows that $\bar{H}^{r}_{x}$  is equivalent to $(\bar{C}^{r}_{x})^{n}.$  A vector in $(\bar{C}^{r}_{x})^{n}$ corresponds to an n-tuple, $\{a^{r}_{x,j}:j=1,\cdots ,n\}$ of number values in $\bar{C}^{r}_{x}.$ Use of the fact that the value $a_{x,j}^{r}$ in $\bar{C}^{r}_{x},$ corresponds to the number value $ra_{j,x}$ in $\bar{C}_{x}$ shows that  the n-tuple in $(\bar{C}^{r}_{x})^{n}$ corresponds to the n-tuple, $r\{a_{j,x}:j=1,\cdots,n\}$ in $\bar{C}_{x}^{n}.$ These equivalences should extend to the case where $\bar{H}_{y}$ and $\bar{H}_{x}$ are separable, which is the case here.} Here $r\psi_{x}$ is the same vector in $\bar{H}^{r}_{x}$ as $\psi_{x}$ is in $\bar{H}_{x}.$

 The description of $\bar{H}^{r}_{x}$ given here is suitable for use in section \ref{QT} where wave packets  for quantum systems are discussed.  For gauge theories,  the Hilbert spaces contain vectors for the internal variables of matter fields. In this case one has to include a gauge field to account for the freedom to choose bases \cite{Montvay,Yang}. The local representation of $\bar{H}_{y}$ on $\bar{H}_{x}$ is then given by \cite{BenIJTP,BenSPIE}
 \begin{equation}\label{barrVHx}\bar{H}^{r,V}_{x}= \{H_{x},\pm_{x},\frac{\cdot_{x}}{r}, \frac{\langle -,-\rangle_{x}}{r},rV\psi_{x}\}.\end{equation} If the $\bar{H}_{x}$ are $n$ dimensional, then $V$ is an element of the gauge group, $U(n).$

 \section{Gauge fields}\label{GF}

 As was noted in the introduction, for $y=x+\hat{\nu}dx,$ $r_{y,x}$ can be represented as the exponential of a vector field:  \begin{equation}\label{cyxGamma}r_{y,x}=e^{\vec{A}(x)\cdot \hat{\nu}dx}=e^{A_{\nu}(x)dx^{\nu}}\end{equation}(sum over repeated indices implied).  $\vec{A}(x)$ is also referred to as a gauge field as it gives the relations between neighboring complex number structures at different space time points. To first order in small quantities, \begin{equation}\label{1storderGx}
 r_{y,x} =1+\vec{A}(x)\cdot \hat{\nu}dx.\end{equation}

 The use of $r_{y,x}$ makes  clear the fact that the setup described here is a generalization of the usual one.  To see this,  set $\vec{A}(x)=0$ everywhere.  Then $r_{y,x}=1$ for all $y,x$ and the local representations of $\bar{C}_{y}$ and $\bar{R}_{y}$ on $\bar{C}_{x}$ and $\bar{R}_{x}$ are $\bar{C}_{x}$ and $\bar{R}_{x}.$ Since the $\bar{C}_{x}$ and $\bar{R}_{x}$ are then independent of $x$,  one can replace $\bar{C}_{x}$  and $\bar{R}_{x}$ with just one complex and real number structure, $\bar{C}$ and $\bar{R}.$

 \subsection{Scale factors for distant points}

 The description of $r_{y,x}$ can be extended to points $y$ distant from $x.$ Let $P$ be a path from $x$ to $y$ parameterized by a real number, $s,$ such that $P(0)=x$ and $P(1)=y.$ Let $r_{y,x}^{P}$ be the scale factor associated with the path $P$.  If $a_{y}$ is a number value in $\bar{C}_{y}$, then $a_{y}$ corresponds to the number value, $r^{P}_{y,x}a_{x},$ in $\bar{C}_{x}$ where $a_{x}=F_{x,y}a_{y}$ is the same number value in $\bar{C}_{x}$ as $a_{y}$ is in $\bar{C}_{y}.$

 One would like to express $r^{P}_{y,x}$ as an exponential of a line integral along $P$ of  the field $\vec{A}(x).$ However this is problematic because the integral corresponds to a sum over $s$ of complex number values in $\bar{C}_{P(s)}.$ Such a sum is not defined because addition is defined only within a  number structure.  It is not defined between different structures.

 This can be remedied by referring all terms in the sum to one number structure such as $\bar{C}_{x}$.  To see how this works, consider a two step path from $x$ to $y=x+\hat{\nu_{1}}\Delta_{x}$ and from $y$ to $z=y+\hat{\nu_{2}}\Delta_{y}.$ $\Delta_{y}$ is the same number in $\bar{C}_{y}$ as $\Delta_{x}$ is in $\bar{C}_{x}.$

 Let $a_{z}$ be a number value in $\bar{C}_{z}.$ $a_{z}$ corresponds to the number value $r_{z,y}\times_{y}a_{y}$ in $\bar{C}_{y}.$ Here $a_{y} =F_{y,z}a_{z}$ is the same number value in $\bar{C}_{y}$ as $a_{z}$ is in $\bar{C}_{z}.$ In $\bar{C}_{x},$ $r_{z,y}\times_{y}a_{y}$ corresponds to the number value  given by \begin{equation}\label{cyxczy}
 r_{y,x}\times_{x}(r_{z,y})_{x}(\frac{\times_{x}}{r_{y,x}}) (r_{y,x}\times_{x}a_{x})=(r_{z,y})_{x}r_{y,x}a_{x}.\end{equation} Here $(r_{z,y})_{x}=F_{x,y}(r_{z,y})$ is the same number in $\bar{C}_{x}$ as $r_{z,y}$ is in $\bar{C}_{y}$  and $\times_{x}/r_{y,x}$ is the representation of $\times_{y}$ on $\bar{C}_{x}.$ The $\bar{C}_{x}$ multiplications are implied in the righthand term and $a_{x}=F_{x,y}a_{y}.$

 The factor $(r_{z,y})_{x}r_{y,x}$ can be expressed in terms of the field $\vec{A}.$ It is \begin{equation}\label{expGyGx}
 (r_{z,y})_{x}r_{y,x}=e^{(\vec{A}(y))_{x}\cdot\hat{\nu_{2}} \Delta_{x}+\vec{A}(x)\cdot\hat{\nu_{1}} \Delta_{x}}.\end{equation} $\Delta_{y}$ is replaced here by its same value $\Delta_{x}$ in $\bar{C}_{x}.$

    Let $P$ be an $n$ step path where $P(0)=x_{0}=x,
    P(j)=x_{j}, P(n-1)=x_{n-1}=y$ and
    $x_{j+1}=x_{j}+\hat{\nu}_{j}\Delta_{x_{j}}.$  Then
    $ r^{P}_{y,x}a_{x}$ is given by\begin{equation}\label{cPyx}
    r^{P}_{y,x}=\prod_{j=0}^{n-1}(r_{x_{j+1},x_{j}})_{x}
    =\exp(\sum_{j=0}^{n-1}[(\vec{A}(x_{j}))
    \cdot\hat{\nu}_{j}\Delta_{x_{j}}]_{x}).\end{equation} The
    subscript $x$ denotes the fact that all terms in the product
    and in the exponential, are values in
    $\bar{C}_{x}.$  For example $(r_{x_{j+1},x_{j}})_{x}=F_{x,x_{j}} r_{x_{j+1},x_{j}}$ is the same value in $\bar{C}_{x}$ as $r_{x_{j+1},x_{j}}$ is in $\bar{C}_{x_{j}}.$  An ordering of terms in the product of Eq,
    \ref{cPyx} is not needed because the different $r$
    factors commute with one another.

 This can be extended  to a line integral along $P$.  The result is \cite{BenSPIE}\begin{equation}\label{cPyxCx}
 r^{P}_{y,x}=\exp\{\int_{0}^{1}(\vec{A}(P(s)))_{x}\cdot(\frac{dP(s)}
 {ds})_{x}ds\}=\exp\{\int_{P}(\vec{A}(\vec{z}))_{x}d\vec{z}\}.\end{equation} The subscript $x$ on the factors in the integral mean that the terms are all evaluated in $\bar{C}_{x}.$

 It is unknown if the field $\vec{A}$ and thereby $r^{P}_{y,x}$ is or is not independent of the path $P$ from $x$ to $y.$  If $\vec{A}$ is not integrable, then the path dependence introduces complications.  In particular it means that for $y$ distant from $x,$ there is no path independent way to describe the local representation of $\bar{C}_{y}$ on $\bar{C}_{x}.$ The local representation would have to include a path variable as in $\bar{C}^{r^{P}_{y,x}}_{x}.$

 In this work, this complication will be avoided by assuming that $\vec{A}$ is integrable. Then $r^{p}_{y,z}=r_{y,x},$ independent of $P$.

 Let $P$ be a path from $x$ to $y$ and $Q$ be another path from $y$ to $x$. Then integrability gives \begin{equation}\label{QP}r^{Q*P}_{x,x}=(r^{Q}_{x,y})_{x} r^{P}_{y,x}=1_{x}\end{equation} Here $Q*P$ is the concatenation of $Q$ to $P$. This result gives \begin{equation}\label{Pinv}(r^{Q}_{x,y})_{x} =(r^{P}_{y,x})^{-1}.\end{equation}

 \section{Local availability of mathematics}

  The local availability of mathematics means that for an observer, $O_{x},$ at point $x$, the mathematics that $O_{x}$ can use, or is aware of, is locally available at $x$. Since mathematical systems are represented by structures, \cite{Barwise,Keisler},  one can use $\bigvee_{x}$  to denote the collection of all these structures.  $\bigvee_{x}$ includes real and complex number and Hilbert space  structures $\bar{R}_{x},\bar{C}_{x},$ $\bar{H}_{x},$ structures for operator algebras as well as many other structure types. All the mathematics that $O_{x}$ uses to make  physical predictions and  physical theories use the systems in $\bigvee_{x}.$  Similarly, all the mathematics available to an observer $O_{y}$ at point $y$ is contained in $\bigvee_{y}.$

 An important requirement is that the mathematics available, in principle at least,  to an observer must be independent of the observers location.  This means that $\bigvee_{y}$ must be equivalent to $\bigvee_{x}$.  For each system structure  in $\bigvee_{y},$ there must be a corresponding structure in $\bigvee_{x}.$  Conversely, for each system structure in $\bigvee_{x}$ there must be a corresponding structure in $\bigvee_{y}.$ Furthermore the corresponding structures in $\bigvee_{y}$ and $\bigvee_{x}$ must be related by parallel transforms that map one structure to another.   These parallel transforms define what is meant by the \emph{same} structure and the same structure elements and operations in $\bigvee_{x}$ as in $\bigvee_{y}$, and conversely.

 This use of parallel transforms is an extension to other types of mathematical systems, of the definitions and use of parallel transforms, Section \ref{FNS}, to relate complex and real number structures at different space time points. It is based on the description of each type of mathematical system as structures, each consisting of a base set, basic operations, relations and constants, that satisfy a set of axioms relevant to the system type.\cite{Barwise,Keisler}.

 The association of an observer to a point, as in $O_{x},$ is an idealization,  mainly because observers, e.g. humans, have a finite size.  Because of this, an observer's location is a region and not a point. This is the case  if one notes that the observer's brain is the seat of all mathematical knowledge and limits consideration to  the brain.  In addition, quantum mechanics introduces an inherent uncertainty to the location of any system. In spite of these caveats, the association of an observer to a point will be used here.

 An important aspect of  $\bigvee_{x}$ is that $O_{x}$ must be able to use the systems in $\bigvee_{x}$ to describe the systems in $\bigvee_{y}.$ This can be done by means of parallel transform maps or correspondence maps from systems in $\bigvee_{y}$ to those in $\bigvee_{x}.$ Parallel transforms map elements and operations of system structures $\bar{S}_{y}$ to the same elements and operations of $\bar{S}_{x}.$ In this case $O_{x}$ can use the mathematics of $\bar{S}_{x}$ as a stand in for the mathematics of $\bar{S}_{y}.$

 Correspondence maps take account of scaling of real and complex numbers in relating systems at $y$ to those at $x.$  In this case $O_{x}$ describes the mathematics of $\bar{S}_{y}$ in terms of the local representation, $\bar{S}^{r}_{x},$  of $\bar{S}_{y}$ on $\bar{S}_{x}.$ If $S=R$ or $S=C$ then $O_{x}$ would describe the properties of $\bar{R}_{y}$ or $\bar{C}_{y}$ in terms of the local scaled systems $\bar{R}^{r}_{x}$ and $\bar{C}^{r}_{x}.$

 The existence of correspondence maps means that for each system type, $S,$ $\bigvee_{x}$ contains all the scaled systems, $\bar{S}^{r}_{x},$ for each point $y$, in addition to $\bar{S}_{x}.$ (Recall $r=r_{y,x}.$)  They include scaled real numbers, $\bar{R}^{r}_{x},$ complex numbers, $\bar{C}^{r}_{x},$ and scaled Hilbert spaces, $\bar{H}^{r}_{x},$ as well as many other system types.

 All these scaled systems are available to an observer, $O_{x}$ at $x$. Since they are locally available, $O_{x}$ takes account of scaling by using them to make theoretical calculations that require inclusion of numbers or vectors at different space or space time points. If $O_{x}$ does not use these correspondence maps and restricts use to parallel transform maps only, then the setup becomes simpler in that each $\bar{S}^{r}_{x}$ is identical to $\bar{S}_{x}.$

 This raises the question of when correspondence maps can be used instead of parallel transform maps.  This will be discussed in the next sections. Here the use of correspondence maps follows from the inclusion into physics of the freedom to choose number systems at different space time points. In this sense it extends the freedom to choose bases in vector spaces in gauge field theory \cite{Yang,Montvay} to the underlying scalars.

 \section{Correspondence maps and parallel transform maps}\label{CMPTM}

  It is proposed here that correspondence maps be used in any theoretical physics expression that requires the mathematical comparison of mathematical entities at different space time points.  Typical examples are space and time derivatives  or integrals of functions or functionals  as maps from space time (or space and time) to elements of mathematical systems that are based on the real or complex numbers as scalars. An example is an $n$ component complex scalar field.

 At this point it is not completely clear if there are other cases in which correspondence maps should be used instead of parallel transform maps. However it is clear that there are many situations where these maps should not be used. These involve what is referred to here as the commerce of mathematics and physics.\footnote{Mathematical and physical commerce refers to the use of numerical values as outcomes of computations and experiments in science, business, and communication.}

 To see this suppose $O_{x}$ wants to compare the numerical output of either an experiment or a theoretical computation, done at $x,$  with the numerical output of either an experiment or computation done at $y.$ Let $b_{x}$ and $d_{y}$ be the real valued numerical outcomes obtained. Use of the correspondence maps means that $O_{x}$ would compare $b_{x}$ with the local representation of $d_{y}$ at $x$, that is, with the number $r_{y,x}d_{x}.$ Here $d_{x}$ is the same number in $\bar{R}_{x}$ as $d_{y}$ is in $\bar{R}_{y}.$

  This is  contradicted by experience.  There is no hint of a factor $r_{y,x}$ in comparing outcomes of repeated experiments, or comparing experimental outcomes with theoretical predictions, or in any other use of numbers in commerce.  If one ignores statistical and quantum uncertainties, numerical outcomes of repeated experiments or repeated computations are the same irrespective of when and where they are done.\footnote{A similar criticism of a suggestion by Weyl was made almost 100 years ago by Einstein \cite{OR}.}

 The reason for this is a consequence of a basic fact.  This is that no experiment and no computation \emph{ever} directly yields a numerical value as an outcome. Instead the outcome of any experiment is a physical system in a physical state that is \emph{interpreted} as a numerical value.  Similarly the outcome of a computation is a physical system in a state that is \emph{interpreted} as a numerical value.

 The crucial word here is \emph{interpreted}. If $\psi_{y}$ is the output state of a measurement apparatus for an experiment at point $y,$ and $\phi_{x}$ is the output state of a computation at point $x$, then the numerical values of these output states are given by $a_{y}=I_{y}(\psi_{y})$ and $b_{x}=I_{x}(\phi_{x}).$ Here $I_{y}$ and $I_{x}$ are interpretive maps from the output states of the measurement system into $\bar{R}_{y}$ and from the computation output states into $\bar{R}_{x}$ respectively. The space time dependence of the maps is indicated by the $x,y$ subscripts.

 The "Naheinformationsprinzip", no information at a distance principle \cite{Mack,Montvay}, forbids direct comparison of the information in $\psi_{y}$ with that in $\phi_{x}.$ This means that $I_{y}(\psi_{y})$ and $I_{x}(\phi_{x})$ cannot be directly compared.  Instead the information contained in $\psi_{y}$ and that contained in $\phi_{x}$ must be transported, by physical means,  to a common point  for comparison.

 There are many different methods of physical information transmission. Included are optical and electronic methods as well as older slower methods.  All methods involve motion of an information carrying physical system from one point to another, "information is physical" \cite{Landauer}.  The physical system used should be such that the state of the information carrying degrees of freedom does not change during transmission from one point to another.

\begin{figure}[h!]\begin{center}
 \rotatebox{270}{\resizebox{150pt}{150pt}{\includegraphics[80pt,130pt]
 [490pt,540pt]{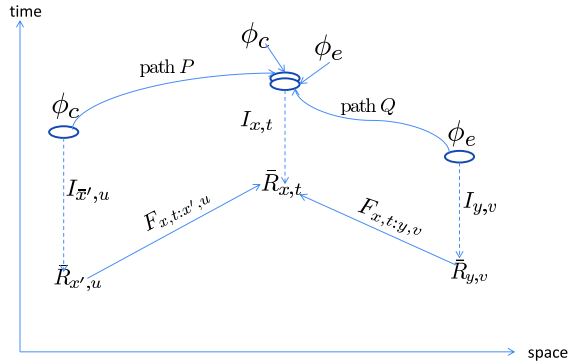}}}\end{center}
 \caption{A simple example of comparing theory with experiment.   The ovals denote the output computation and experiment systems in states $\phi_{c}$ and $\phi_{e}$ at  space and times, $x',u$ and $y,v.$ $P$ and $Q$ denote the paths followed  by these systems. One has $P(u)=x'$ and $Q(v)=y.$ The double oval in the center denotes the the two systems at the point, $x$, of path intersection where $P(t)=Q(t)=x.$  The interpretation maps are denoted by $I_{P(s),s}$ and $I_{Q(s),s}$ for different times $s$.  The real number structures $\bar{R}_{P(s),s}$ and $\bar{R}_{Q(s),s}$ are associated with each point in the paths $P$ and $Q.$  $F_{x,t;x',u}$ and $F_{x,t:y,v}$ are parallel transform operators that map the real number structures at the points of theory and experiment completion to the point of path intersection.}\label{LAM1}\end{figure}

 Figure \ref{LAM1} illustrates schematically,  in one dimensional space  and time, the nonrelativistic transmission  of a theory computation output state  obtained at $x',u$ and an experimental output state obtained at $y,v$  to a common point, $x,t,$ for comparison. Here $x,x',y$ are space locations and $u,v,t$ are times.

 The figure, and the discussion, illustrate a general principle. All activities in the commerce of mathematics and physics consist of physical procedures and operations that generate physical output systems in states that are interpreted as numerical values. The "no information at a distance" principle forbids direct comparison of the associated number values at different points.  Instead the systems or suitable information carriers must be brought to a common point where the numerical information, as number values in just one real number structure, can be locally compared.

 Similar considerations apply to storage of outcomes of experiments or computations either in external systems or in the observers brain. As physical dynamic systems, observers move in space time. If $P$ is a path taken by an observer,  with $P(\tau)$ the observers location at proper time $\tau,$ the mathematics available to $O_{P(\tau)}$ is that in $\bigvee_{P(\tau)}.$ If $\phi(P(\tau))$ denotes the state of a  real number memory trace in an observers brain, then the number value represented by $\phi(P(\tau))$ is given by  $I_{P(\tau)}(\phi(P(\tau))).$ This is a number value in $\bar{R}_{P(\tau)}.$ At a later proper time $\tau^{\prime},$ the number value represented by the memory trace is $I_{P(\tau^{\prime})}(\phi(P(\tau^{\prime})).$ If there is no degradation of the memory trace, then $I_{P(\tau)}(\phi(P(\tau)))$ is the same number value in $\bar{R}_{P(\tau)}$ as $I_{P(\tau^{\prime})} (\phi(P(\tau^{\prime}))$  is in $\bar{R}_{P(\tau^{\prime})}.$ Correspondence maps play no role here either.

  \section{Quantum theory}\label{QT}
   As might be expected, the local availability of mathematics and the freedom of choice of number  scaling factors, have an effect on quantum theory.  This is a consequence of the use of space time integrals and derivatives in the theory.  For example,  one would expect to see the gauge field $\vec{A}$ appear in quantum descriptions of physical systems.  To see how this effect arises, it is useful to limit the treatment to  nonrelativistic quantum mechanics on three dimensional Euclidean space, $R^{3}.$

   \subsection{Effect of the local availability of mathematics on quantum theory}\label{LAM}
   The local availability of mathematics requires that the usual setup of just one $\bar{C},\bar{R},\bar{H}$ is replaced by separate number structures, $\bar{R}_{x},\bar{C}_{x}$ and separate Hilbert spaces, $\bar{H}_{x},$ associated with each $x$ in $R^{3}.$\footnote{The association of separate Hilbert spaces to each point $x$ is different here from that used in gauge theory \cite{Montvay}. In gauge theory, the spaces are all finite dimensional and apply to internal states of the Fermion fields.  Here the Hilbert spaces describe states of systems spread over space, e.g. as wave packets.} It follows that mathematical operations, such as space or time derivatives or integrals, which involve nonlocal mathematical operations on numbers or vectors at different points, cannot be done. The reason is that these operations  violate mathematical locality.

   To preserve locality, one must use either parallel transformations or correspondence transformations. These two methods are well illustrated by considering a single particle wave packets.  The usual representation has the form \begin{equation}\label{psiint}\psi =\int\psi(y)|y\rangle dy\end{equation}where the integral is over all space points in $R^{3}.$

   One result of the local availability of mathematics is that, for each $y,$ the vector. $\psi(y)|y\rangle.$ is in $\bar{H}_{y},$ just as $\psi(y)$ is a number value in $\bar{C}_{y}.$  It follows that the space integral over $y$  makes no sense.  It describes a suitable limit of  adding vectors that belong to different Hilbert  spaces.  Addition is not defined between different spaces; it is defined only within one Hilbert space and complex number structure.

    The use of parallel transformations replaces Eq. \ref{psiint} by \begin{equation}\label{psiintx}\psi_{x}= \int_{x}\psi(y)_{x}|y_{x}\rangle_{x} dy_{x}.\end{equation} Here $\psi(y)_{x}=F_{x,y}\psi(y)$ is the same number value in $\bar{C}_{x}$ as $\psi(y)$ is in $\bar{C}_{y},$ the number triple,  $y_{x},$ in $|y_{x}\rangle_{x}$ is the same triple in $\bar{R}^{3}_{x}$ as $y$ is in $\bar{R}^{3}_{y}$, and $|y_{x}\rangle_{x}$  is the same state in $\bar{H}_{x}$ as $|y\rangle$ is in $\bar{H}_{y}.$ The differentials $dy_{x}=dy^{1}_{x}dy^{2}_{x}dy^{3}_{x}$ refer to $\bar{R}^{3}_{x}.$   The subscript $x$ in $\int$ indicates that the integral is based on $\bar{H}_{x},\bar{C}_{x}.$ The representations of sameness given above are shown explicitly by \begin{equation}\label{Fxypsi}
   \begin{array}{c}\psi(y)_{x}=F_{x,y}\psi(y)\\y_{x}=
   \bar{F}_{x,y}(y)\\|y_{x}\rangle_{x}=|\bar{F}_{x,y}(y) \rangle_{x}\\dy_{x}=\bar{F}_{x,y}(dy).\end{array}\end{equation}   Also $|\bar{F}_{x,y}(y) \rangle_{x}$ is the same basis vector in $\bar{H}_{x}$ as $|y\rangle$ is in $\bar{H}_{y}.$

    Note that the point $x$ on which the integral is based is arbitrary.   Eq. \ref{psiintx} holds if the subscript $x$ is replaced by another point $z$. Then the integral is based on $\bar{C}_{z},\bar{H}_{z}.$

    The use of parallel transforms is applicable to other aspects of quantum mechanics. For  each $y$ in $R^{3},$ the momentum operator, $\textbf{p}_{y},$ for vectors in $\bar{H}_{y}$ is given by\begin{equation}\label{ihdel}\textbf{p}_{y} =i_{y}\hbar_{y}\nabla_{y}=i_{y}h_{y} \sum_{j=1}^{3}\partial_{j,y}.\end{equation}Here, $i_{y},\hbar_{y}$ are numbers in $\bar{C}_{y}.$  The action of $\textbf{p}_{y}$ on a vector $\psi$  at point $y$ gives for the $jth$ component\begin{equation}\label{dely}\textbf{p}_{y,j} \psi=i_{y}\hbar_{y}\partial_{j,y}\psi =\frac{\psi(y+dy^{j})-\psi(y)}{dy^{j}}. \end{equation} As was the case for the space integral, this expression makes no sense because $\psi(y+dy^{j})$  is in $\bar{C}_{y+d^{j}y}$ and $\psi(y)$ is in $\bar{C}_{y}.$

   This can be remedied by replacing $\partial_{j,y}$ by $\partial'_{j,y}$ where \begin{equation}\label{delz}\partial'_{j,y}\psi =\frac{\psi(y+dy^{j})_{y}-\psi(y)}{dy^{j}}.\end{equation} Here $\psi(y+dy^{j})_{y}=F_{y,y+d^{j}y}\psi(y+dy^{j}).$ It follows from this that the expression for the momentum becomes \begin{equation}\label{boldppy} \textbf{p}'_{y}\psi =\sum_{j=1}^{3}p'_{j,y}\psi=i_{y}\hbar_{y} \sum_{j=1}^{3}\partial'_{j,y}\psi.\end{equation}

    The Hamiltonian for  a single quantum system in an external potential, acting on a state, $\psi_{y}$ at point $y$, is given by \begin{equation}\label{Hamy} H_{y}\psi(y) =-\frac{\hbar^{2}_{y}} {2m_{y}}\sum_{j=1}^{3} (\partial')^{2}_{y,j} \psi(y)+V(y) \psi(y).\end{equation}  Here, $\hbar_{y}$ and $m_{y}$ are Planck's constant and the particle mass. They have values in $\bar{R}_{y}.$ The values of the external potential, $V(y),$ are also in $\bar{R}_{y}.$

   The main difference between this and the usual expression for a Hamiltonian is the replacement of $\partial_{j,y}$ with $\partial'_{j,y}.$  Otherwise, the expressions are the same.

   For a single particle state, $\psi,$ the momentum representation is $\psi=\int\psi(p)|p\rangle dp.$ Here $dp=dp_{1}dp_{2}dp_{3}.$ Since the amplitude $\psi(p)$ is a complex number value and no location for the value is specified, one may choose any location, $x,$ such as that of an observer, $O_{x},$ to assign $\psi(p)$ as a number value in $\bar{C}_{x}$ and the integral as an element of $\bigvee_{x}.$

   The relation between $\psi(p)$ and $\psi(x)$ is given by the Fourier transform. The components of the space integral in \begin{equation}\label{FTpsi}\psi(p)=\int e^{i_{z}p_{z}z} \psi(z)dz\end{equation} must all be mapped to a common point, $x,$ for the integral to make sense. This gives \begin{equation}\label{FTpsiz} \psi(p)_{x}=\int_{x} (e^{i_{z}p_{z}z})_{x}\psi(z)_{x}dz_{x}\end{equation} Here \begin{equation}\label{eipxz} (e^{i_{z}p_{z}z})_{x}=F_{x,z}e^{i_{z} p_{z}z}=e^{i_{x}p_{x}z_{x}}\end{equation}is the same number in $\bar{C}_{x}$ as $e^{i_{z}p_{z}z}$ is in $\bar{C}_{z}.$

   The treatment described can  be extended to multiparticle  entangled states. It is sufficient to consider two particle states. For example a two particle state $\psi_{1,2}$ where the total momentum of the two particles is $0$ can be expressed by \begin{equation}\label{psipo}\psi_{1,2} =\int\psi_{1}(p)\psi_{2}(-p) |p\rangle_{1}|-p\rangle_{2}dp.\end{equation} Use of Fourier transforms gives \begin{equation}\label{psi12}\psi_{1,2}=\int dz_{1}dz_{2}(\int e^{iz_{1}p}\psi_{1}(p)|p\rangle_{1}e^{-iz_{2}p} \psi_{2}(-p)|-p\rangle_{2}dp).\end{equation}

   The integral must be transformed to a Hilbert space with just one scalar field.  For a point $x$ with $\bar{C}_{x},\bar{H}_{x},$ the integrand factors are parallel transformed to obtain \begin{equation}\label{psi12z}(\psi_{1,2})_{x}=\int_{x} (dz_{1})_{x}(dz_{2})_{x}(\int (e^{iz_{1}p})_{x}\psi_{1}(p)_{x} |p_{x}\rangle_{1}(e^{-iz_{2}p})_{x}\psi_{2}(-p)_{x}|-p_{x}\rangle_{2}dp_{x}).
   \end{equation} Here \begin{equation}\label{ex1zx2z}
   \begin{array}{c}(e^{iz_{1}p})_{x}=F_{x,z_{1}}(e^{iz_{1}p})\\
   (e^{-iz_{2}p})_{x}=F_{x,z_{2}}(e^{-iz_{2}p})\end{array}
   \end{equation}Also $p_{x}$ is the same value in $\bar{C}_{x}$ as $p$ is in $\bar{C}_{z_{1}}$ in the $z_{1}$ integral, Eq. \ref{psi12}, and $-p_{x}$ is the same value in $\bar{C}_{x}$ as $-p$ is in $\bar{C}_{z_{2}}$ in the $z_{2}$ integral.

   \subsection{Inclusion of number system scale factors}\label{INSSF}

   The above shows that the imposition of "mathematics is local" on quantum theory  is more complex than the usual treatment  with just one scalar field and one Hilbert space  for all space points. Since the description with parallel transforms is equivalent to the usual one, the added complexity is not needed if one goes no further with it.

  This is not the case if one extends the treatment to include space dependent scaling factors for the different $\bar{C}_{x},\bar{R}_{x}.$   For a given $x$, the local representations of $\bar{C}_{y}$ on $\bar{C}_{x}$ are given by scaled representations, $\bar{C}^{r_{y,x}}_{x},$ of $\bar{C}_{y}$ on $\bar{C}_{x}.$ Also the local representation of $\bar{H}_{y}$ on $\bar{H}_{x}$ with effects of the number scaling included, is given by $\bar{H}^{r_{y,x}}_{x}.$

   For $y=x+\hat{\nu}dx,$ a neighbor point of $x,$ the scaling factor, $r_{y,x}$ is given by, $r_{y,x}=e^{\vec{A}(x)\cdot\hat{\nu}dx},$ Eq. \ref{cyxGamma}.  If $y$ is distant from $x$ and $\vec{A}(x)$ is integrable, then, expressing $r_{y,x}$ as an integral along a straight line path from $x$ to $y$ gives, (Eq. \ref{cPyxCx})\begin{equation}\label{ryxstrline}r_{y,x} =\exp(\int_{0}^{1}\sum_{i=1}^{3}A_{i}(sx_{i})_{x}s(x_{i})_{x}ds)= \exp(\sum_{i=1}^{3} \int_{x_{i}}^{y_{i}}A_{i}(z_{i})_{x}dz^{i}_{x}) =\prod_{i=1}^{3}r_{y,x,i}\end{equation} Here $x_{i}=\vec{x}\cdot\hat{i}$ and $y_{i}=\vec{y}\cdot\hat{i}$ are the components of $\vec{x}$ and $\vec{y}$ in the direction $i.$ The last equality assumes that the components of $\vec{A}$ commute with one another.\footnote{Here, and in what follows, $\vec{A}(x)$ is assumed  to be integrable.} The subscript $x$ indicates that the integrals are  defined on $\bar{R}_{x}.$

   The presence of $\vec{A}(x)$ affects the expression of a wave packet state $\psi$ as given by  Eq. \ref{psiintx}.  In this case the wave packet expansion of $\psi_{x}$ is  given by \begin{equation}\label{psiintAFxy} \psi_{x}=\int_{x}r_{y,x}\psi(y)_{x}|y_{x}\rangle dy_{x}\end{equation} where $r_{y,x}$ is given by Eq. \ref{ryxstrline}.

   This result is obtained by noting that for each $y$ the local representation of $\bar{H}_{y},\bar{C}_{y}$ on $\bar{H}_{x},\bar{C}_{x},$ with scaling factor included, is $\bar{H}^{r_{y,x}}_{x},\bar{C}^{r_{y,x}}_{x}.$ The vector in $\bar{H}^{r_{y,x}}_{x}$ that is the same vector as $\psi(y)|y\rangle$ is in $\bar{H}_{y},$ is denoted by $\psi(y)^{r}_{x}\cdot^{r}_{x}|y^{r}_{x}\rangle.$ Here $\psi(y)^{r}_{x}$ is the same number value in $\bar{C}^{r_{y,x}}_{x}$ as $\psi(y)$ is in $\bar{C}_{y}$ and $|y^{r}_{x}\rangle$ is the same vector in $\bar{H}^{r_{y,x}}_{x}$ as $|y\rangle$ is in $\bar{H}_{y}.$ This follows from the observation that $y^{r}_{x}$ and $y$ are the space positions  associated with $|y^{r}_{x}\rangle$ and $|y\rangle$ in $\bar{H}^{r_{y,x}}_{x}$ and $\bar{H}_{y}$ respectively.  Scalar vector multiplication in $\bar{H}^{r_{y,x}}_{x}$ is shown by $\cdot^{r}_{x}.$

   The corresponding state on $\bar{H}_{x}$ is obtained by noting that \begin{equation}\label{psirx}
   \begin{array}{l}\psi^{r}_{x}\cdot^{r}_{x}|y^{r}_{x}\rangle\Rightarrow r_{y,x}\psi(y)_{x}(\cdot^{r}_{x})_{x}|r_{y,x}y_{x}\rangle=
   r_{y,x}\psi(y)_{x}\frac{\cdot_{x}}{r_{y,x}}r_{y,x}|y_{x}\rangle
   =r_{y,x}\cdot_{x}|y_{x}\rangle.\end{array}\end{equation}This is the result shown in Eq. \ref{psiintAFxy}. The use of $(\cdot^{r}_{x})_{x}=\frac{\cdot_{x}}{r_{y,x}},$ Eq. \ref{barHrx}, is based on the requirement that $\bar{H}^{r_{y,x}}_{x}$ satisfies the same Hilbert space axioms \cite{Kadison} as does $\bar{H}_{x}.$

   It must be emphasized that the usual predictions of the quantum mechanical properties of wave packets, with $\vec{A}(x)=0$ everywhere, do a good job of prediction of experimental results. So far, quantum mechanical predictions and experiments have not shown the need for the presence of $\vec{A}.$ This shows that the effect of the $\vec{A}$ field must be very small, either through the values of the field itself or by use of a very small coupling constant, $g$, of the field to numbers and vectors. This would be accounted for by replacing $\vec{A}$ in Eqs. \ref{cyxGamma} and \ref{ryxstrline} by $g\vec{A}.$

   In this sense the presence of $\vec{A}$ is no different than the presence of the gravitational field. In theory, a proper description of quantum mechanics of systems should include the effects of the gravitational field.  However, it can be safely neglected because the field is so small, at least far away from black holes where quantum physics is done.

   Another  feature of Eq. \ref{psiintAFxy} is the dependence on the reference point $x$. This can be removed by restricting the integration volume to a region of space, excluding $x,$ where the region  contains essentially all of $\psi.$ This is what one does in any experiment since $\psi$ is prepared in a region that does not include the observer.

   The removal of $x$ dependence then follows from expressing  Eq. \ref{psiintAFxy}  as a sum of two terms, one as the integral over $V$ and the other over all space outside $V$:\begin{equation}\label{psiintAVWFxy} \psi_{x}=\int_{x,V}r_{y,x}\psi(y)_{x}|y_{x}\rangle dy_{x}+\int_{x,W}r_{y,x}\psi(y)_{x}|y_{x}\rangle dy_{x}.\end{equation} The subscript, $W,$ on the second integral means that it refers to integration over all space outside $V.$ Here $x$ is a point in $W$.

   Assume that $V$ is chosen so that the integral over $W$ can be neglected. Then \begin{equation}\label{psiintAVFxy} \psi_{x}\cong \int_{x,V}r_{y,x}\psi(y)_{x}|y_{x}\rangle dy_{x}.\end{equation} This equation has a problem in that the correspondence transforms are extended  from any point in $V$ to a point outside $V.$ However these transforms are restricted here to apply within space or space time integrals, and not outside the integration volume.

\begin{figure}[h!]\begin{center}
 \rotatebox{270}{\resizebox{120pt}{120pt}{\includegraphics[80pt,130pt]
 [490pt,540pt]{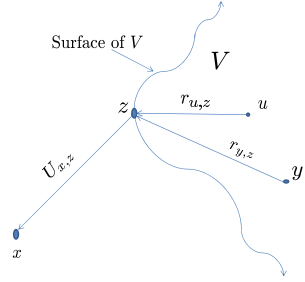}}}\end{center}
 \caption{Representation of scaling factors  in the integrals from point $z$ on the surface of $V$  to points $y$ and $u$. The direction implied in the order of the subscripts of $r$ is opposite to the direction of the parallel transformations taking the integrand factors from $y$ and $u$ to $z.$ $U_{x,z}$ parallel transforms $\psi_{z}$ to $\psi_{x,z}$.  This is the same vector in $\bar{H}_{x}$ as $\psi_{z}$ is in $\bar{H}_{z}.$}\label{LAM4}\end{figure}

   This can be fixed by choice of a point $z$ on the surface of $V$  and replacing $r_{y,x}$ by $U_{x,z}r_{y,z}.$  The factor $r_{y,z}$ accounts for the correspondence transform from a point $y$ in $V$ to a point $z$ on the surface of $V$, and $U_{x,z}$ is a unitary operator that parallel transforms the result from $z$ to $x.$ Then one has \begin{equation}\label{psiintAVFxy} \psi_{x,z}\cong U_{x,z}\psi_{z}= U_{x,z}\int_{z,V}r_{y,z}\psi(y)_{z}|y_{z}\rangle dy_{z}=
   \int_{x,V}(r_{y,z})_{x}\psi(y)_{x}|y_{x}\rangle dy_{x}.\end{equation} The subscript $z$ on $\psi_{x,z}$ indicates a possible dependence on the choice of $z$ on the surface of $V$. Figure \ref{LAM4} illustrates the setup for two points $y,u$ in $V$.

   This result shows that the wave packet representation is independent of $x$, such as an observers location, provided it is outside of $V$.  However, to the extent that $\vec{A}$ cannot be neglected, the representation does depend on the location of $z$.  To see what this dependence is, let $w$ be another point on the surface of $V$. Then, following Eq. \ref{psiintAVFxy}, $\psi_{x,w}$ is given by \begin{equation}\label{psiintAVFxyw} \psi_{x,w}\cong U_{x,w}\psi_{w}= U_{x,w}\int_{w,V}r_{y,w}\psi(y)_{w}|y_{w}\rangle dy_{w}=
   \int_{x,V}(r_{y,w})_{x}\psi(y)_{x}|y_{x}\rangle dy_{x}.\end{equation}

   For the comparison, at $x$, of $\psi_{x,z}$ with $\psi_{x.w},$  it is sufficient to compare, in $\bar{H}_{z},$  the parallel transformation of $\psi_{w}$ to $z$ with $\psi_{z},$ in $\bar{H}_{z}.$ The parallel transformation of $\psi_{w}$ to $z$ is given by \begin{equation}\label{psiwzy}(\psi_{w})_{z}= \int_{z,V}(r_{y,w})_{z}(\psi(y)_{w})_{z} |(y_{w})_{z}\rangle d(y_{w})_{z}.\end{equation}Use of the fact that parallel transforms of numbers and and vectors from $y$ to $w$ and then to $z$ are the same as transforms from $y$ to $z$ gives \begin{equation}\label{psiwzy1}(\psi_{w})_{z}= \int_{z,V}(r_{y,w})_{z}\psi(y)_{z} |y_{z}\rangle dy_{z}.\end{equation}

   Since $\vec{A}$ is integrable, one can write
   \begin{equation}\label{rywyz}(r_{y,w})_{z}=(r_{y,z}r_{z,w})_{z} =r_{y,z}(r_{z,w})_{z}\end{equation} to obtain \begin{equation}\label{psiwzy2}(\psi_{w})_{z}=(r_{z,w})_{z} \int_{z,V}r_{y,z}\psi(y)_{z} |y_{z}\rangle dy_{z}.\end{equation} The subscript, $z,$ denotes parallel transformation to $z$, of mathematical elements that are not  in $\bigvee_{z}.$ No subscript appears on $r_{y,z}$ as it is already a number value in $\bar{R}_{z}.$

   This shows that $(\psi_{w})_{z}$ differs from $\psi_{z}$ by a factor, $(r_{z,w})_{z}=(r_{w,z})^{-1},$ Eq. \ref{Pinv}. The difference is preserved on parallel transformation to $x$ in that $(\psi_{w})_{x}=\psi_{x,w}$ differs from $(\psi_{z})_{x}=\psi_{x,z}$ by a factor $(r_{z,w})_{x}.$

   If the effect of $\vec{A}$ is small, then it is useful to express $r_{y,x}$ as an expansion to first order in the exponential. For example the expression for $\psi_{x}$ in Eq. \ref{psiintAFxy} becomes, \begin{equation}\label{psicong} \psi_{x}\cong \int_{x}(1+\int_{x}^{y} \vec{A}(w)_{x}\cdot\hat{\nu}dw_{x}) \psi(y)_{x}|y_{x}\rangle dy_{x}.\end{equation} Here $\hat{\nu}$ is a  unit vector along the direction from $x$ to $y.$ The first term of the expansion corresponds to the usual case with $\vec{A}$ equal to $0$ everywhere. The $x$ dependence arises from the second term, which gives the correction due to the presence of $\vec{A}.$

    The restriction of the integration to a finite volume $V,$ as in Eq. \ref{psiintAVFxy}, removes the dependence on $x$ in that $\psi_{x}$ is the same vector in $\bar{H}_{x}$ as $\psi_{y}$ is in $\bar{H}_{y}$ provided $y$ is not in $V$. Expansion of $(r_{y,z})_{x}$ to first order in small terms shows that the $z$ dependence arises from the $\vec{A}$ containing term as in Eq. \ref{psicong}.

    The dependence on $z$ can be appreciable because $z$ can be any point on the surface of $V$.   What is interesting is that this dependence can be greatly reduced by using the properties of actual measurements to minimize the effect of $\vec{A}$ on the predicted expectation value.

   Consider a position measurement  on a system in state $\psi_{x},$ Eq. \ref{psiintAFxy}. The expectation value for this measurement, calculated at $x,$ is given by
   \begin{equation}\label{popsix}\langle \psi_{x}|\tilde{y}|\psi_{x}\rangle_{x}= \int_{x}r_{y,x}y_{x}|\psi(y)|^{2}_{x}dy_{x}.\end{equation}This expectation value\footnote{All but one of the $r$ factors appearing in the integrand are canceled by the $r$ factors in the denominators of the multiplication operations.}  is an idealization or what one does. It does not take account of what one actually does.

   Typically, position measurements are done by dividing a  volume of space up into cubes and measuring the relative frequency of occurrence of the quantum system in the different cubes.  A measurement consists of a large number of repetitions of this measurement on repeated preparations of the system in state $\psi.$ Assume the cubes in space have volume $\Delta^{3}$ where $\Delta$ is the length of a side. Then outcomes of the repeated experiment are "yeses" from the cube detectors whose locations are denoted by  triples, $j,k,l$ of integers. Each "yes" means that the location is somewhere in the volume of the responding detector cube located at position, $z_{j,k,l}=j\Delta, k\Delta, l\Delta.$ The local availability of mathematics means that $z_{j,k,l}$ is a triple of numbers in $\bar{R}_{z_{j,k,l}}.$

   The presence of parallel and correspondence transformations enables physical theory to express exactly what is done experimentally. Eq. \ref{psiintAFxy} for $\psi_{x}$ is replaced by an expression that limits integrals with scaling factors to the cube volumes and parallel transports these integrals to a common point $x$ where they can be added together.  The result, $\psi'_{x},$ is given by\begin{equation}\label{psijklsum}
   \psi'_{x} =\sum_{j,k,l}U_{x,z}\int_{V_{j,k,l}}r_{w,z}\psi(w)_{z}|w_{z}\rangle dw_{z}.
   \end{equation}Here the sum is over all cubes. Each cube is labeled by a point $z=z_{j,k,l}$ on the cube surface. Each integral  over the volume, $V_{j,k,l}=\Delta^{3},$ of cube, $j,k,l$, is a vector in $\bar{H}_{z}.$ Within each integral, the $r$ factor scales the values of each integrand at point $w$ to values at $z.$  $U_{x,z}$ parallel transforms the integrals at  different $z$ to a common point $x.$

   The theoretical expectation value for the experimental setup described here is given by \begin{equation}\label{exppos}
   \langle\psi'|\tilde{y}|\psi'\rangle_{x}= \sum_{j,k,l}U_{x,z} \int_{V_{j,k,l}}r_{w,z}w_{z}|\psi(w)|^{2}_{z}dw_{z}.\end{equation} The effect of the $r$ factor is  smaller here than it is in the expectation value using $\psi_{x}.$ The reason is that it is limited to integrations over small volumes.

   This representation of the prediction is supported by  the discussion on mathematical and physical commerce. The "no information at a distance" principle requires that the information contained in the outcomes of each position measurement, as physical systems in "yes" or "no" states for each point $z_{j,k,l},$  be transmitted by physical means to $x$ where the results of the repeated measurements are tabulated. The tabulation is all done at $x.$  No factor involving $\vec{A}$ appears in the transmission or tabulation.

   As noted, the effect of $\vec{A}$ appears only in the integrals over the volumes $\Delta^{3}.$ In these integrals, to first order,\begin{equation}\label{rzw}r_{w,z}\cong 1_{z}+\int_{z}^{w}\vec{A}(y)_{z} \cdot\hat{\mu}dy_{z}.\end{equation} Since the integral is limited to points within the volume $\Delta^{3},$ it is clear that as $\Delta\rightarrow 0,$ the integrals for each cube also approach $0$.

   This shows that the effect of the $\vec{A}$ field diminishes as the accuracy of the measurement increases. In the limit  $\Delta=0,$ the $\vec{A}$ field disappears and one gets the usual theoretical prediction without $\vec{A}$ present. However, the Heisenberg uncertainty principle prevents the limit, $\Delta =0,$  from actually being achieved.

   The presence of the $\vec{A}$ field affects other quantum mechanical properties of systems. For example, the description of the momentum operator with $\vec{A}\neq 0,$ replaces Eq. \ref{boldppy}  by \begin{equation}\label{boldAppy} \textbf{p}_{A,y}\psi =\sum_{j=1}^{3}p_{A,j,y}\psi=i_{y}\hbar_{y} \sum_{j=1}^{3}D_{j,y}\psi.\end{equation}$D_{j,y}\psi$ is given by altering Eq. \ref{delz} to read \begin{equation}\label{Delyz} D_{j,y}\psi =\frac{r_{y+dy^{j},y}\psi(y+dy^{j})_{y}-\psi(y)} {dy^{j}}.\end{equation}Here $r_{y+dy^{j},z}\psi(y+dy^{j})_{y}$ is the number value in $\bar{C}_{y}$ that corresponds to $\psi(y+dy^{j})$ in $\bar{C}_{y+dy^{j}}.$

   Using the fact that \begin{equation}\label{rzdjz}r_{y+d^{j}y,y}= e^{A_{j}(y)d^{j}y}\end{equation} and expansion to first order in the exponential gives, \begin{equation}\label{DApsi}D_{j,y}\psi=\partial'_{j,y} \psi+A_{j}(y)\psi(y).\end{equation} The momentum components become
   \begin{equation}\label{canonp}p_{A,j,y}=i_{y}\hbar_{y}D_{j,y}=i_{y}\hbar_{y}
   (\partial'_{j,y}+A_{j}(y)).\end{equation}This expression is similar to that for the canonical momentum in the presence of an electromagnetic field.  Note that $A_{j}(y)$ is pure real.

   The expressions for the Hamiltonian for a single particle remain as shown in Eq. \ref{Hamy}  except that $\partial'$ is replaced by $D.$ For example Eq. \ref{Hamy} becomes\begin{equation}\label{HamAyx}(H_{y})\psi(y) =-\frac{\hbar^{2}_{y}} {2m_{y}}\sum_{j=1}^{3} (D)^{2}_{j,y} \psi(y)+V(y) \psi(y)\end{equation} with $D_{y,j}$ given by Eq. \ref{DApsi}.

   Inclusion of  scaling factors into the two particle state entangled by momentum conservation is straightforward. This is achieved by including scale factors in the two particle space integral, $\int_{x} (dz_{1})_{x}(dz_{2})_{x},$ in Eq. \ref{psi12z}.  The result is \begin{equation}\label{psi12Az}\begin{array}{l}(\psi_{1,2})_{x}=\int_{x}r_{z_{1},x} (dz_{1})_{x}r_{z_{2},x}(dz_{2})_{x}\\\\ \hspace{1cm}\times (\int (e^{iz_{1}p})_{x}\psi_{1}(p)_{x} |p_{x}\rangle_{1}(e^{-iz_{2}p})_{x}\psi_{2}(-p)_{x}|-p_{x}\rangle_{2}dp_{x}).\end{array}
   \end{equation}Here $r_{z_{1},x}$ and $r_{z_{2},x}$ are given by Eq. \ref{ryxstrline}.

 \section{Gauge theories}
 One approach \cite{Montvay} to gauge theories already makes partial use of the local availability of mathematics with the assignment of an $n$ dimensional vector space to each $x$. Here the vector space is assumed to be a Hilbert space, $\bar{H}_{x},$ at each $x.$ This $\bar{H}_{x}$ is quite different from that discussed in the previous section in that the vectors in $\bar{H}_{x}$ refer to the internal states of matter fields. Matter fields $\psi$  are functionals where for each space time point $x,$ $\psi(x)$ is a vector in $\bar{H}_{x}.$

 The freedom of choice of a basis \cite{Yang,Montvay} in each $\bar{H}_{x}$ is reflected in the factorization, \begin{equation}\label{UYV}U_{y,x}= Y_{y,x}V_{y,x},\end{equation}of a parallel transform operator, $U_{y,x},$ \cite{Mack} from $\bar{H}_{x}$ to $\bar{H}_{y}$ where $y=x+\hat{\nu}dx$ is a neighbor point of $x.$\footnote{Factorization is necessary because $U_{y,x}$ cannot be expressed as an exponential of Lie algebra elements or as a matrix of numbers. The reason is that the action of such a representation on a vector in $\bar{H}_{x}$ gives another vector in $\bar{H}_{x}.$ It is not a vector in $\bar{H}_{y}$.  Factorization remedies this in that $V_{y,x}$ is a unitary map from $\bar{H}_{x}$ to $\bar{H}_{x}$ and $Y_{y,x}$ is a unitary map from $\bar{H}_{x}$ to $\bar{H}_{y}.$}

 The unitary operator $V_{y,x}$ expresses the freedom of basis choice. As such it is an element of the gauge group $U(n)$ with a Lie algebra representation \cite{Utiyama,Cheng} \begin{equation}\label{LieVyx}
 V_{y,x}=e^{i\Xi_{\mu}(x)dx^{\mu}}e^{i\Omega^{j}_{\mu}(x)\tau_{j}dx^{\mu}}. \end{equation}Sum over repeated indices is implied. The $\tau_{j}$ are the generators of the Lie algebra $su(n)$ and the $\Omega^{j}_{\mu}(x)$ are the components of the $n$ different gauge fields, $\vec{\Omega}^{j}(x).$ $\vec{\Xi}(x)$ is the gauge field for the $U(1)$ factor of $U(n).$

 The covariant derivative of the field, $\psi,$ is expressed by
 \begin{equation}\label{Dmuxreg}D_{\mu,x}\psi=\frac{V_{\mu,x} \psi(x+dx^{\mu})_{x}-\psi(x)}{dx^{\mu}}.\end{equation}Here $V_{\mu,x}$ is the $\mu$ component of $V_{y,x}.$ Expansion of the exponential to first order in small quantities gives\begin{equation}\label{Dmuxreg1}D_{\mu,x}\psi =\partial^{\prime}_{\mu,x}\psi+i(g_{1}\Xi_{\mu}(x)+g_{2}
 \Omega^{j}_{\mu}(x)\tau_{j})\psi(x).\end{equation} Coupling constants, $g_{1}$ and $g_{2},$ have been added.  The definition of $\partial^{\prime} _{\mu,x}\psi$  is  essentially the same as that given in Eq. \ref{delz}. It is given by\begin{equation}\label{partial}\partial^{\prime}_{\mu,x}\psi= \frac{\psi(x+dx^{\mu})_{x}-\psi(x)}{dx^{\mu}}.\end{equation} Here $\psi(x+dx^{\mu})_{x}=U_{x,x+dx^{\mu}}\psi(x+dx^{\mu})$ is the same vector in $\bar{H}_{x}$ as $\psi(x+dx^{\mu})$ is in $\bar{H}_{x+dx^{\mu}}.$

 The covariant derivative, Eq. \ref{Dmuxreg1}, accounts for the local availability of mathematics and the freedom of basis choice.  It does not include the effects of scaling factors for numbers. This is taken care of by replacing $V_{\mu,x} \psi(x+dx^{\mu})_{x}$ in Eq. \ref{Dmuxreg} by $r_{x+dz^{\mu},x}V_{\mu,x} \psi(x+dx^{\mu})_{x}$.  This is a vector in the local representation, $\bar{H}^{r}_{x},$ Eq. \ref{barHrx}, of $\bar{H}_{y}$ on $\bar{H}_{x}.$

 Expansion of the exponentials to first order adds another term to $D_{\mu,x}$ in Eq. \ref{Dmuxreg1}. One obtains \cite{BenIJTP,BenSPIE}\begin{equation}\label{Dmuxreg2}D_{\mu,x}\psi =\partial^{\prime}_{\mu,x}\psi+g_{r}A_{\mu}(x)+i(g_{1}\Xi_{\mu}(x)+g_{2}
 \Omega^{j}_{\mu}(x)\tau_{j})\psi(x).\end{equation} A coupling constant, $g_{r},$ for $\vec{A}(x)$ has been added.  The coupling constants, and $i$ are all number values in $\bar{C}_{x}.$

 The physical properties of the gauge fields in $D_{\mu,x}$ are obtained by restricting the Lagrangians to only those terms that are invariant under local and global gauge transformations \cite{Cheng}. For Abelian gauge theories, such as QED, $\vec{\Omega}(x)$ is absent.    Invariance under local gauge transformations, $\Lambda(x)$, requires that  the covariant derivative satisfy \cite{Cheng}\begin{equation}\label{DpD}D'_{\mu,x}\Lambda(x)\psi(x) =\Lambda(x)D_{\mu,x}\psi(x).\end{equation}$D'_{\mu,x}$ is obtained from $D_{\mu,x}$ by replacing $A_{\mu}(x)$ and $\Xi_{\mu}(x)$ with their primed values, $A'_{\mu}(x),\Xi'_{\mu}(x).$ The presence of the primes allows for the possible dependence of the fields on the local  $U(1)$ gauge transformation, $\Lambda(x)$ where \begin{equation}\label{Lamx}\Lambda(x) =e^{i\phi(x)}.\end{equation}

 Use of Eq. \ref{partial} and separate treatment of real and imaginary terms gives the following results: \cite{Cheng}
 \begin{equation}\label{ApABpB}\begin{array}{c}A'_{\mu}(x)=A_{\mu}(x) \\\ g_{1}\Xi'_{\mu}(x)=g_{1}\Xi_{\mu}(x) -\partial'_{\mu,x}\phi(x).\end{array}\end{equation} This shows that the real field $\vec{A}$ is unaffected by a $U(1)$ gauge transformation. It also shows that $\Xi_{\mu}(x)$ transforms in the expected way as the electromagnetic field.

 As is well known the properties of the $\vec{\Xi}$ field show that it is massless.  The reason is that a mass term for this field is not locally gauge invariant \cite{Montvay,Cheng}.

 Unlike the case for the  $\vec{\Xi}$  field, a mass term can be present for the real $\vec{A}$ field.  This suggests that it represents a gauge boson for which mass is optional.  That is, depending on what physical system $\vec{A}$ represents, if any, the presence of a mass term  in Lagrangians is not forbidden.

 For nonabelian gauge theories, such as $U(2)$ theories, Eq. \ref{ApABpB} still holds.  However there is an additional equation giving the transformation properties of the three vector gauge fields under local $SU(2)$ gauge transformations. These properties result in the physical representation of these fields in Lagrangians as charged vector bosons \cite{Cheng}. The  $\vec{A}$ and $\vec{\Xi}$ bosons  are still present.

 \subsection{Physical properties of the $\vec{A}$  field from the gauge theory viewpoint}
 At this point it is not known what physical system, if any, is represented by the $\vec{A}$ field. Candidates include the inflaton field \cite{Linde,Albrecht}, the Higgs boson, the graviton, dark matter, and dark energy. One aspect that one can be pretty sure of is that the ratio of the $\vec{A}$ field -  matter field coupling constant, $g_{r}$, to the fine structure constant, $\alpha,$ must be very small. This is a consequence of the great accuracy of the QED Lagrangian and the fact that the $\vec{A}$ field appears in  covariant derivatives for all gauge theory (and other) Lagrangians.

  As was noted, $\vec{\Xi}$ is the photon field. Inclusion of this  field and a Yang Mills term for the dynamics of this field into the Dirac Lagrangian gives the QED Lagrangian \cite{Cheng}.

 \section{Conclusion}
 This work is based on two premises: the local availability of mathematics and the existence of scaling factors for number systems. Local availability is based on the idea that the only mathematics that is directly available to an observer is that which is, or can be, in his or her head. Mathematical information that is separate from an observer, $O_{x},$ at space time point $x$, such as a textbook or a lecturer at point $y,$ must be physically transmitted, e.g. by acoustic or light waves, to $O_{x}$ where it becomes directly available.

 This leads to a setup in which mathematical universes, $\bigvee_{x},$ are associated with each point $x.$ If an observer moves through space time on a world line, $P(\tau),$ parameterized by the proper time $\tau,$ the mathematics directly available to $O_{P(\tau)}$ at time $\tau$ is that in $\bigvee_{P(\tau)}.$

 Each $\bigvee_{x}$ contains many types of mathematical systems. If $\bar{S}_{x}$ is in $\bigvee_{x}$, then $\bigvee_{y}$ contains the same system type, $\bar{S}_{y},$ and conversely. Each $\bigvee_{x}$ contains the different types of number systems and many other systems that are based on numbers. Included are the real and complex numbers $\bar{R}_{x}$ and $\bar{C}_{x}.$

 Here the mathematical logical definition \cite{Barwise,Keisler} of each type of system as a structure is used. A structure consists of a base set, basic operations, relations, and constants that satisfies axioms appropriate for the type of structure considered. Examples are $\bar{R}$ and $\bar{C}$, Eq. \ref{barRC} for the real and complex numbers.

 For each type of number structure it is possible to define many structures of the same type that differ by scaling factors \cite{BenRENT}. For each real number $r$, one can define structures $\bar{R}^{r}$, Eq. \ref{barRrr}, and $\bar{C}^{r}$, Eq. \ref{barCcc}, in which a scale factor $r$ relates the number values in $\bar{R}^{r}$ and $\bar{C}^{r}$  to those in $\bar{R}$ and $\bar{C}.$  The scaling of number values must be compensated for by scaling of the basic operations and relations in a manner such that $\bar{R}^{r}$ and $\bar{C}^{r}$ satisfy the relevant axioms for real and complex numbers if and only if $\bar{R}$ and $\bar{C}$ do.

 The local availability of mathematics requires that one be able to construct local representations of $\bar{C}_{y}$ on $\bar{C}_{x}.$ Two methods were described. One uses parallel transformations. These define or represent the notion of sameness between mathematical systems at different points.  If $F_{x,y}$ is a parallel transform map from $\bar{S}_{y}$ onto $\bar{S}_{x},$ then for each element, $w_{y}$, in $\bar{S}_{y},$  $w_{x}=F_{x,y}(w_{y})$ is the same element in $\bar{S}_{x}$ as $w_{y}$ is in $\bar{S}_{y}.$  In this case the local representation, $W_{x,y}\bar{S}_{y},$ of $\bar{S}_{y}$ on $\bar{S}_{x}$ is $\bar{S}_{x}$  itself.

 The other method uses what are called correspondence maps. These combine parallel transformations with scaling. The local representation of $\bar{C}_{y}$ on $\bar{C}_{x}$ is $\bar{C}^{r}_{x},$ which is a scaling of $\bar{C}_{x}$ by a factor $r=r_{y,x}.$ (From now on $\bar{R}$  is not explicitly mentioned as it is implicitly assumed to be part of $\bar{C}.$) The local representation of an element, $a_{y},$ of $\bar{C}_{y}$  corresponds to the element $r_{y,x}a_{x}$ in $\bar{C}_{x}.$ Here $a_{x}=F_{y,x}a_{y}$ is the same element in $\bar{C}_{x}$ as $a_{y}$ is in $\bar{C}_{y}.$

 It was seen that the scaling of numbers plays no role in the general use of numbers in mathematics and physics. This includes such things as comparing outcomes of theory predictions with experimental results or in comparing outcomes of different experiments.  More generally it plays no role in the use of numbers in the commerce of mathematics and physics. The reason is that theory computations and experiment outcomes obtained at different locations are never  directly compared.  Instead the information contained in the outcomes  as physical states must be transmitted  by physical systems  to a common point. There the states of the physical transmittal systems are interpreted locally as numbers, and then compared.

 In this work, number scaling was limited to theory calculations that involve space time derivatives or integrals. Examples of this were described in quantum theory and in gauge theories. An example discussed in some detail was the expansion of a wave packet $\psi=\int\psi(y)|y\rangle dy.$ Since $\psi(y)|y\rangle$ is a vector in $\bar{H}_{y},$ the integrand has to be moved to a common point, $x$ for the integral to make sense. This can be done by parallel transform maps which give \begin{equation}\label{psipt}
 \psi_{x}=\int_{x}\psi(y)_{x}|y_{x}\rangle dy_{x}\end{equation} or by correspondence maps which give \begin{equation}\label{psirpt}\psi_{x}=\int_{x}r_{y,x}\psi(y)_{x}|y_{x}\rangle dy_{x}.\end{equation} Here the scaling factor, $r_{y,x},$ is the integral from $x$ to $y$ of the exponential of the gauge field, $\vec{A}(y),$ as in Eq. \ref{ryxstrline}.

 It was also seen that one can use both transform and correspondence maps to express the wave packet in a form that reflects exactly what one does in an experiment that measures either the spatial distribution or the position expectation value of a quantum particle. If the experiment setup consists of a collection of cube detectors of volume $\Delta^{3}$ that fill $3$ dimensional Euclidean space, the outcome of each of many repeated experiments is a triple of numbers, $j,k,l$ that label the position, $j\Delta,k\Delta.l\Delta$ of the detector that fired.

 As was seen in the discussion of mathematical and physical commerce, the outcomes of repeated experiments must be physically transported to a common point, $x$ where they are interpreted as numbers in $\bar{R}_{x}$ for mathematical combination. It follows that the scaling factors are limited to integration over the volume of each detector.  This results in the replacement of $\psi_{x}$ by $\psi'_{x}$ where, Eq. \ref{psijklsum}, \begin{equation}\label{psijklsum1}\psi'_{x} =\sum_{j,k,l}U_{x,z} \int_{V_{j,k,l}}r_{w,z}\psi(w)_{z}|w_{z}\rangle dw_{z}.
   \end{equation}The sum is over all cubes. $z=z_{j,k,l}$ is a point on each cube surface. Each integral is over the volume, $V_{j,k,l},$ of each cube. The $r$ factor, Eq. \ref{rzw}, scales the values of each integrand at point $w$ to values at $z.$  $U_{x,z}$ parallel transforms the integrals at  different $z$ to a common point $x.$

 The state, $\psi'_{x},$ differs from $\psi_{x},$ or the usual quantum mechanical wave packet expression for $\psi,$ in that it ties theory closer to experiment. It also reduces the effect of scaling to the sum of the effects for the volumes of each of the detectors. The effect is reduced because, for any point $w$ in the sensitive volume of the experiment, the effect of $\vec{A}$ on the transform from $w$ to $x$ is limited to the part of the path in the detector volume containing $w.$  As the detector volumes go to $0$, so does the effect of $\vec{A}.$ The increase in the number of detectors as the volume of each gets smaller does not remove this effect.

 In this sense the usual quantum theory wave packet integral for $\psi$ is a limit in that it is independent of experimental details.  Unlike the case for the usual representation of $\psi$, use of $\psi'_{x}$ to make predictions  will give values that depend on experimental details. The fact that there is no indication, so far, of such dependence, at least to the accuracy of experiment, means that the effect of $\vec{A}(x)$ must be very small. Whether the small effect is due to small values of $\vec{A}$ itself or a small value of a coupling constant of $\vec{A}$ to states and matter fields, is not known at present.

 The relation between $\psi'_{x}$ and the usual integral for $\psi$ is further clarified by the observation that $\psi_{x}\rightarrow \psi$ as the detector volume goes to $0$.  However, the Heisenberg uncertainty principle prevents experimental attainment of this limit.

 It must be emphasized that the tying the wave packet integral to experiment details, as with $\psi'_{x}$ has nothing to do with collapse of the wave packet during carrying out of the experiment. $\psi'_{x}$ is just as coherent a state as is $\psi.$

 The gauge field $\vec{A}$ also appears in the expression for the canonical momentum.  The usual expression for momentum $\textbf{p}=\sum_{j} i\hbar\partial_{j,x}$ is replaced by, Eq. \ref{boldAppy}
 \begin{equation}\label{pax1}\textbf{p}_{A,x}=i_{x}\hbar_{x}D_{j,x}
 \end{equation} where \begin{equation}\label{canp}D_{j,x}=\partial'_{j,x} +\vec{A}.\end{equation} $\partial'_{j,x}$, Eq. \ref{delz}, accounts for the local availability of mathematics.

 As a covariant derivative, $D_{\mu,x}$ appears in gauge theories  with additional terms.  It was seen that the limitation of Lagrangians to terms that are invariant under local gauge transformations, \cite{Montvay,Cheng}, results in $\vec{A}$ appearing as a gauge boson for which mass is optional. This is the case for both Abelian and nonabelian gauge theories.

 The physical nature of $\vec{A}$, if any, is unknown. What is known is that the great accuracy of QED requires that the coupling constant of $\vec{A}$ to matter fields must be very small.

 It must be emphasized that this work is only a first step in  combining "mathematics is local"  with the freedom of choice of scaling factors for number structures.  An example of work for the future is to determine the effect of number scaling factors on geometry. It is suspected that the scaling factors may induce conformal transformations into geometry.  More work also needs to be done on the effects of number scaling on quantum mechanics. An interesting question here is whether scaling factors are needed at all in classical mechanics.

 Finally, one may hope that this work provides a real entry into the description of a coherent theory of physics and mathematics together. Such a theory would be expected to describe mathematics and physics together as part of a coherent whole instead of as two separate but closely related disciplines.

            \section*{Acknowledgement}
          This work was supported by the U.S. Department of Energy,
          Office of Nuclear Physics, under Contract No.
          DE-AC02-06CH11357.


\begin{thebibliography}{99}

              \bibitem{Wigner}
            Wigner, E. (1960). The unreasonable effectiveness of mathematics in the natural sciences, \textit{Commum. Pure and Applied Math.} {\bf 13}, 001
            (1960), Reprinted in E. Wigner, {\it Symmetries and Reflections},
            (Indiana Univ. Press, Bloomington IN, pp. 222-237.

             \bibitem{Bernal}
            Bernal, A.; Sanchez, M.; Gil, F. (2008). Physics ftrom scratch. Letter on M. Tegmark's, 'The mathematical universe", arXiv:0803.0944.


            \bibitem{Bendaniel}
           BenDaniel, D. (1999). Linking the foundations of physics and mathematics, arXiv:math-phy/9907004.


            \bibitem{Davies}
            Davies, P. (1990). Why is the Physical World so Comprehensible? in
            \textit{Complexity, Entropy, and Physical Information},
            Proceedings of the 1988 workshop on complexity, entropy, and the
            physics of information, may-june 1989, Santa Fe New Mexico, W. H.
            Zurek, Editor, Addison-Weseley Publishing Co. Redwood City CA,
            61-70.

            \bibitem{Hut}
            Hut,P.; Alford, M.; Tegmark, M. (2006). On math, matter, and mind,
            \emph{Foundations of Physics}, \textbf{36}, 765-794, 2006;
            arXiv: physics/ 0510188.

            \bibitem{Jannes}
             Jannes, G. (2009) Some comments on "The mathematical universe", \emph{Foundations of Physics}, \textbf{39}, 397-406, 2009.

              \bibitem{Omnes}
           Omnes, R. (2011). Wigner's "Unreasonable Effectiveness of
           Mathematics", Revisited, \emph{Foundations of Physics},
           \textbf{41}, 1729-1739, 2011.

            \bibitem{Tegmark}
            Tegmark, M.  (2008). The mathematical universe, \emph{Foundations of Physics}, \textbf{38}, 101-150, 2008; arXiv:0704.0646.

            \bibitem{Welch}
            Welch, L. (2009). A possible mathematical structure for physics, arXiv:0908.2063.

             \bibitem{BenTCTPMTEC}
             Benioff, P. (2005). Towards a coherent theory of  physics and mathematics: the theory experiment connection, \emph{Foundations of Physics},\textbf{35}, 1825-1856, 2005;  arXiv:quant-ph/0403209.


            \bibitem{BenTCTPM}
            Benioff, P. (2002). Towards a coherent theory of physics and mathematics, \emph{Foundations of Physics},\textbf{32}, 989-1029, 2002;
            arXiv:quant-ph/0201093.

            \bibitem{Barwise}
             Barwise, J. (1977). An Introduction to First Order Logic, in
            \emph{Handbook of Mathematical Logic}, J. Barwise, Ed.
            North-Holland Publishing Co. New York,  5-46.

             \bibitem{Keisler}
            Keisler, H. (1977). Fundamentals of Model Theory, in
            \emph{Handbook of Mathematical Logic}, J. Barwise, Ed.
            North-Holland Publishing Co. New York, 1977, 47-104.





              \bibitem{Kaye}
            Kaye, R. (1991). \emph{Models of Peano Arithmetic} Clarendon Press, Oxford,
            1991,  16-21.


            \bibitem{real}
            Randolph, J. (1968). \emph{Basic Real and Abstract Analysis},
            Academic Press, Inc. New York, NY,  p 26.

             \bibitem{complex}
            Shoenfield, J. (1967). \emph{Mathematical Logic}, Addison Weseley
            Publishing Co. Inc. Reading Ma, p. 86.


            \bibitem{Kadison}
           Kadison, R. \&  Ringrose, J. (1983). \emph{Fundamentals of
           the Theory of Operator Algebras: Elementary theory},
           Academic Press, New York,  Chap 2.


             \bibitem{BenIJTP}
            Benioff, P. (2011). New gauge field from extension of
            space time parallel transport of
            vector spaces to the underlying number systems, \emph{International Journal of theoretical physics} \textbf{50}, 1887-1907 2011;  arXiv:1008.3134

            \bibitem{BenRENT}
            Benioff, P.  (2011).  Representations of each number type that differ by scale factors, arXiv:1102.3658


               \bibitem{BenSPIE}
            Benioff, P. (2011). New gauge fields from extension of parallel transport of vector spaces to underlying scalar fields, \emph{Proceedings of SPIE, Quantum Information and Computation IX} \textbf{8057},80570X, Orlando, Florida, USA, April, 2011, SPIE, Bellingham, Washington.

              \bibitem{Yang}
            Yang , C. \& Mills, R. (1954). Conservation of isotopic spin and isotopic gauge invariance, \emph{Physical Review} \textbf{96}, 191-195, 1954.


              \bibitem{Montvay}
            Montvay, I. \&  M\"{u}nster, G. (1994). \emph{Quantum Fields on a Lattice}, Cambrodge University Press, Cambridge, UK,  Chapter 3.



            \bibitem{Novaes}
           Novaes, S. (2000). Standard model: an Introduction, in
           \emph{Particles and Fields}, Proceedings, X
           Jorge Andre Swieca Summer School, Sao Paulo,
           February 1999, Editors, J. Barata, M. Begalli,
           R. Rosenfeld, World Scientific Publishing Co.
           Singapore; arXiv:hep-th/0001283.


            \bibitem{Mack}
            Mack, G. (1981). Physical principles , geometrical aspects, and locality properties of gauge field theories, \emph{Fortschritte der Physik}, \textbf{29}, 135-185, 1981.

            \bibitem{Landauer}
            Landauer, R. (1991). Information is Physical, \emph{Physics Today}, \textbf{44}, 23-29, 1991.

              \bibitem{OR}
            O'Raifeartaigh, L. (1997). \emph{The Dawning of Gauge Theory},
            Princeton Series in Physics, Princeton University
            Press, Princeton, N. J.

               \bibitem{Utiyama}
           Utiyama, R. (1956). Invariant Theoretical Interpretation of Interaction, \emph{Phys. Rev.} \textbf{101}, 1597-1607, 1956.


                 \bibitem{Cheng}
            Cheng, T. \&  Li, L. (1984). \emph{Gauge Theory of Elementary
            Particle Physics}, Oxford University Press, New York, NY,
             Chapter 8.

                 \bibitem{Linde}
           Linde, A. (1982). A new inflationary universe scenario: A possible solution of the horizon, flatness, homogeneity, isotropy and primordial monopole problems, \emph{Phys. Letters B}, \textbf{108}, 389-393, 1982.

             \bibitem{Albrecht}
          Albrecht, A. \&  Steinhardt, P. (1982). Cosmology for Grand Unified Theories with Radiatively Induced Symmetry Breaking, \emph{Phys. Rev. Lett.}
           \textbf{48}, 1220-1223.

            \end{thebibliography}
            \end{document}